\documentclass[12pt,letterpaper]{article}
\pdfoutput=1

\usepackage{graphicx}
\usepackage[body={17.5cm, 22cm},right=2cm]{geometry}
\usepackage{amssymb}
\usepackage{amsmath} 
\usepackage{semtrans}
\usepackage{slashed}
\usepackage{bbold}

\newcommand\ee{\end{equation}}
\newcommand\be{\begin{equation}}
\newcommand\eea{\end{eqnarray}}
\newcommand\bea{\begin{eqnarray}}
\newcommand{\sfrac}[2]{{\textstyle\frac{#1}{#2}}}
\newcommand\di{\partial}

\usepackage{hyperref}
\usepackage{verbatim}
\usepackage[usenames,dvipsnames]{xcolor}
\usepackage[numbers]{natbib}
\usepackage[normalem]{ulem}

\numberwithin{equation}{section}

\begin{document}

\begin{center}

\hfill Imperial/TP/2020/LA/01\\
\vskip 1 cm

{\Large\bf Spontaneously broken boosts\\[.3cm]
and the Goldstone continuum}
\end{center}

\bigskip

\begin{center}

{\large Lasma Alberte$^{a}$,
 Alberto Nicolis$^{b}$
} 

\vskip 0.3cm

{\em $^a$Theoretical Physics, Blackett Laboratory, Imperial College, London, SW7 2AZ, U.K.} \\
{\em $^b$Center for Theoretical Physics and Physics Department, Columbia University, \\New York, NY 10027, USA}

\end{center}

\bigskip

\centerline{\bf Abstract}

\smallskip

\noindent
The spontaneous breaking of boost invariance is ubiquitous in nature, yet the associated Goldstone bosons are nowhere to be seen. We discuss why some subtleties are to be expected in the Goldstone phenomenon for spontaneously broken boosts, and derive the corresponding quantum mechanical, non-perturbative Goldstone theorem. Despite similarities with more standard Goldstone theorems, we show by examples that ours can be obeyed by quite unusual spectra of low-energy excitations. In particular, for non-relativistic Fermi liquids, we prove that it is obeyed by the {\em particle-hole continuum}. To the best of our knowledge, this is the first example of a Goldstone theorem obeyed by a continuum rather than by (approximately stable) single-particle Goldstone boson states.

\pagebreak
\tableofcontents

\section{Introduction}
To the best of our knowledge, Lorentz invariance is a fundamental symmetry of nature. Yet, with the exception of the vacuum, all
states of matter we typically consider break boost invariance. For states made up of a finite number of particles, such as those that we consider for scattering experiments, there is a preferred reference frame---the so called center-of-mass frame, where the total momentum vanishes. For states with a finite density of particles per unit volume,  even if we adopt a coarse-grained description that cannot resolve the individual particles and treat the whole system as a homogeneous solid or fluid continuum, there is also a preferred frame---the one in which the medium is at rest.

Such a breaking of boost invariance is of the spontaneous type: the dynamical laws are Lorentz invariant, but the state of the system is not. Goldstone's theorem and its several variations have taught us to expect, in situations like this, gapless excitations \cite{Goldstone:1961eq,Goldstone:1962es}. So, are there gapless Goldstone bosons associated with the spontaneous breaking of boosts?

There happen to be subtleties related to the spontaneous breaking of spacetime symmetries, having to do, in particular, with the relationship between the number of broken generators and the number of independent Goldstone excitations \cite{Nielsen:1975hm,Nicolis:2012vf}. At the level of the effective field theory for the Goldstone fields, such subtleties are associated with the so-called inverse Higgs constraints \cite{Ogievetsky, Ivanov:1975zq}.  These correspond to conditions expressing some Goldstone fields as certain derivatives of other Goldstone fields.
For instance, for superfluids, the phonon field $\pi(x)$ can be thought of as the Goldstone field associated with the spontaneous breaking of a $U(1)$ symmetry, say particle number. On the other hand, the Goldstone fields for boosts need not be independent degrees of freedom, because it turns out that certain non-linear combinations of $\pi(x)$ and its derivatives \cite{Nicolis:2013lma},
\be
\vec \eta(x) \propto \vec \nabla \pi(x) + {\cal O}(\pi^2) \; ,
\ee
have the right transformation properties to serve as boost Goldstone fields. Similarly, for solids, the phonon fields $\vec \pi(x)$ are the Goldstone fields for spontaneously broken translations, and certain non-linear combinations of them and their derivatives,
\be
\vec \eta(x) \propto \dot{ \vec \pi}(x) + {\cal O}(\pi^2) \; ,
\ee
can serve as boost Goldstone fields.

The practical rule that emerges from the coset construction to assess when a phenomenon like this can take place involves only  the symmetry breaking pattern, and is schematically as follows. Calling $\bar P_a$ the set of {\em unbroken translation} generators, $X_A$ the set of {\em broken} generators (for spacetime or internal symmetries), and $\pi_A(x)$ their associated Goldstone fields, if {\em (i)} 
$[\bar P_a, X_1] \supset X_2$, and {\em (ii)} $X_1$ and $X_2$ do not belong to the same irreducible representation of the unbroken symmetries (apart from translations), then it is consistent with all the symmetries---broken and unbroken alike---to express the Goldstone field $\pi_1(x)$ for $X_1$ as a combination of $\pi_2(x)$ and its derivatives:
\be
\pi_1(x) \propto \partial_a \pi_2(x) + {\cal O}(\pi_2 {}^2) \; .
\ee
This is called an inverse Higgs constraint \cite{Ogievetsky, Ivanov:1975zq} (see also, \emph{e.g.} \cite{Nicolis:2013sga} for a more modern discussion).

By now it is clear that, at least for media that are homogeneous on large scales, usually there are no independent Goldstone excitations for boosts, because the Goldstone excitations for other spontaneously broken symmetries can play that role as well.
In particular, based on the general ideas sketched above, Ref.~\cite{Nicolis:2015sra} attempted a general classification of the possible `condensed matter' systems. By these it is meant, there and here, states of matter that spontaneously break the Poincar\'e group down to suitably defined, perhaps suitably coarse-grained time translations, spatial translations, and, possibly, spatial rotations.
Such a classification correctly reproduces the spectrum of low-energy excitations and more in general the infrared dynamics of solids, ordinary fluids, superfluids, and supersolids (see also \cite{Nicolis:2013lma} {for an earlier work}). For these systems, there are always inverse Higgs constraints at work, which remove, in particular, the boost Goldstones as independent degrees of freedom.
Nevertheless, regarding that classification, there are both an oddity and a notable exception, and both have to do with systems that only break boosts and no other symmetries. 

The oddity is the so-called framid: a framid is a hypothetical isotropic and homogeneous medium whose only low-energy degrees of freedom are the Goldstone bosons for spontaneously broken boosts. Its infrared dynamics are described by a consistent effective field theory, but it is admittedly a peculiar system: for instance, contrary to more mundane media, its equilibrium state has the same energy-momentum tensor as a cosmological constant, that is, it has $\langle T^{\mu\nu} \rangle \propto \eta^{\mu\nu}$, oblivious to the fact that Lorentz invariance {\em is} spontaneously broken. Despite being the simplest implementation of the symmetry breaking pattern for condensed matter as defined above, the framid does not seem to be realized in nature as a phase of matter.

The notable exception instead is the degenerate Fermi liquid: it does seem to be realized in nature, it is described by a consistent effective field theory \cite{Polchinski:1992ed}, yet it does not conform to the allegedly `general' classification of \cite{Nicolis:2015sra}. It breaks boosts but no other symmetries, but it cannot be described by the framid's effective field theory, because, for instance, for a non-relativistic Fermi liquid the mass density is much bigger than the other entries of the stress-energy tensor. In particular, $\rho+p\neq 0$ for Fermi liquids. \

In fact, even setting aside  Ref.~\cite{Nicolis:2015sra} and framids, there is a more general paradox concerning Fermi liquids. Where are the boost Goldstones? If no other symmetries are broken apart from boosts, there are no inverse Higgs constraints available to remove the boost Goldstones, which then  should be there. But the low-energy spectrum of bosonic excitations in a Fermi liquid{---corresponding to specific deformations of the Fermi surface---does not provide us with obvious candidates. At zero temperature, there can be zero-sound, which is a spin singlet, and spin waves, which make up a spin triplet. But from Goldstone's theorem one generically expects Goldstone bosons with the same (unbroken) quantum numbers as the broken generators \cite{Weinberg:1996kr}. If boosts are broken but rotations are not, the Goldstone bosons should come in a triplet under rotations. But in the non-relativistic limit spin and orbital angular momentum are independent quantum numbers. So, one expects Goldstones in an orbital triplet/spin singlet representation of rotations (like the broken boost generators), which apparently leaves us only with the $\ell = 1$ zero-sound modes as possible Goldstone candidates. More precisely, at finite momentum, it leaves us with the  helicity zero and helicity $\pm1$ zero-sound modes. However, in Fermi liquid theory such modes have no special status---they are just three out of possibly infinitely many {(or possibly zero)} propagating modes: the existence of any  zero-sound or spin-wave modes in the spectrum depends crucially on the detailed structure of interactions. In fact, in low-temperature experiments on liquid helium-3, zero-sound shows up at best as a broad resonance (with a relative width of more than $20 \%$), and the experimental evidence for spin waves is not as clean \cite{Fak1994, PhysRevB.61.1421}.

Finally, let us mention that here has been a lot of theoretical work on Fermi liquids and many of its properties, including some of the reasons behind the remarkable success of Landau's Fermi liquid theory \cite{Landau:1956zuh, Lifshitz} have been fully or partly understood \cite{Rothstein:2017twg,Rothstein:2017niq, Varma,Polchinski:1992ed}. Also, even more exotic states of matter that are not contained in the classification of \cite{Nicolis:2015sra} have been found in the past decades, like bad metals \cite{Emery:1995zz,Delacretaz:2016ivq}, non-Fermi liquids \cite{Varma} and anisotropic Fermi liquids \cite{Oganesyan:2001sob}, to name a few. In the existing theories describing such materials one is however not really looking at the excitations of the underlying solid that breaks boosts spontaneously. And so, in these theories, boosts are effectively broken explicitly. Understanding these strongly coupled systems often requires going beyond the standard condensed matter methodology and use instead the techniques from the AdS/CFT correspondence (see \emph{e.g.} \cite{Hartnoll:2008vx, Liu:2009dm, Hartnoll:2016apf}). 

Motivated by all of the above, our paper is devoted to understanding the Goldstone phenomenon for spontaneously broken boosts. We start in the next section with a simple argument that shows that, on general grounds, one  {\em should} expect the Goldstone excitations associated with boosts to be peculiar: in general, applying to boosts the standard Goldstone `know-how' does not work. We then prove a non-perturbative Goldstone theorem for boosts. It predicts the existence of gapless excitations, and forces certain matrix elements between such states and the Lorentz breaking state one is perturbing about to be related to certain expectation values. Despite obvious similarities with more standard Goldstone theorems, ours differs from those in an apparently harmless, minor technical way, which however turns out to have far reaching implications. We  check that while for standard systems with Goldstone bosons---framids and superfluids---our theorem is saturated, as expected, by single-particle Goldstone boson states, for a Fermi liquid it is saturated by a {\em particle-hole continuum}. For that system, there are no single-particle Goldstone boson states. What plays that role is instead a continuum in the spectrum, akin to a multi-particle continuum. 

\vspace{.5cm}
\noindent
{\em Notation and conventions:}
We  keep our analysis relativistic all along, but for simplicity we  take the non-relativistic limit at some point when discussing Fermi liquids. We  use the mostly minus metric signature, $+$$-$$-$$-$,  and natural units, $\hbar = c = 1$. When the number of momentum-space delta-functions escalates, we  use a shorthand notation for them: $\delta_{p_1 p_2} \equiv (2 \pi)^3 \delta^3(\vec p_1 - \vec p_2)$. When needed, we  introduce a finite volume $V$ to regulate momentum-space delta-functions: $\delta_{pp} = V$. Finally, for single-particle states we use the so-called non-relativistic normalization: $\langle \vec p_1 | \vec p_2 \rangle = \delta_{p_1p_2}$.

\section{Why boost Goldstones are special}\label{heuristic}

In the next section we will prove a general Goldstone theorem for spontaneously broken boosts, and in the following ones
we will see how different media can obey such a theorem in qualitatively different ways.
Here instead we want to give a somewhat heuristic argument which shows how in the case of spontaneously broken boosts the  Goldstone phenomenon {\em must} feature some peculiarities compared to more standard symmetry breaking patterns.

Consider a zero-temperature medium in a state that, at least at large distances, is homogeneous and isotropic, but breaks boosts. The energy-momentum operator $T^{\mu\nu}(x)$ must have  expectation values of the form
\be \label{vevs}
\langle T^{00} \rangle = \rho \; , \qquad \langle T^{0i} \rangle = 0 \; , \qquad  \langle T^{ij} \rangle = P \,  \delta^{ij} \; ,
\ee
where $\rho$ and $P$ are suitable constants---the energy density and pressure of our state.

Consider now how the total energy and momentum of the system, $P^\mu \equiv \int d^3 x  \, T^{0\mu}$, change if we perform an infinitesimal boost modulated by a weakly space-time dependent rapidity parameter $\vec \eta( x)$ that goes to zero at spatial infinity. To first order in $\vec \eta$ and zeroth order in its derivatives,
\begin{align}
\delta P^0 &  \simeq 2 \int d^3 x  \,  \eta^j(x) \langle T^{0j} \rangle = 0 \\
\delta P^i  &  \simeq  \int d^3 x  \,  \big[ \eta^i(x) \langle T^{00}  \rangle+ \eta^j(x) \langle T^{ij}  \rangle \big] = (\rho+P) \int d^3 x  \,  \eta^i(x)   \; . \label{Pi}
\end{align}
According to the standard relationship between Goldstone particles and Goldstone fields, {\em if} there are Goldstone bosons associated with the spontaneous breaking of boosts, we can think of $\vec \eta(x)$ as the field operator that creates and annihilates them. But then Eq.~\eqref{Pi} creates a potential problem. The reason is that, according to it, the momentum carried by the Goldstones starts linear in $\eta^i(x)$, that is, linear in creation ($a^\dagger$) and annihilation ($a$) operators. But if spatial translations are unbroken, the excitations of the system can be taken to be eigenstates of the momentum operator. This  however is only possible if the momentum operator starts quadratic in the field operators that create and annihilate such excitations, so that there is a chance to have, schematically, $\delta P^i \sim a^\dagger a$.

The only ways out we see are:
\begin{enumerate}

\item 
Spatial translations are in fact also spontaneously broken, and so there is no need for the excitations of the system to be eigenstates of momentum. This is the case relevant for solids. Notice that for solids there is an extra subtlety: not only the boost Goldstones are not momentum eigenstates, they are also not independent excitations, since their field operator $\vec \eta(x)$ can be expressed in terms of the field operator $\vec \pi(x)$  associated with the translation Goldstones (the phonons):
\be
\vec \eta(x) \propto \dot {\vec \pi}(x) + {\cal O}(\pi^2) \; .
\ee 
This last property however has no role in solving the potential problem outlined in the paragraph right after eq.~\eqref{Pi}.
 
\item 
The field operator $\vec \eta(x)$ is in fact the spatial derivative of another local field, in which case the r.h.s.~of \eqref{Pi} integrates to zero. This is the case relevant for superfluids. There, like for solids, the boost Goldstones are not independent excitations, but now the relationship between their field operator and the superfluid phonon field $\pi(x)$ involves a {\em spatial} derivative:
\be
\vec \eta(x) \propto \vec \nabla \pi(x) + {\cal O}(\pi^2) \; .
\ee
 
\item
The prefactor in the r.h.s.~of \eqref{Pi} vanishes:
\be
\rho + P = 0 \; .
\ee 
This is the case relevant for the elusive `framids' \cite{Nicolis:2015sra}.

\end{enumerate}
In particular, there seems to be no room for boost Goldstones in a system such as a non-relativistic Fermi liquid, which spontaneously breaks boosts but no other symmetries, and has a mass density (times $c^2$) much bigger than its pressure.

To proceed, we thus need to understand what the Goldstone theorem actually has to say about the spontaneous breaking of boosts.

\section{A Goldstone theorem for spontaneously broken boosts}\label{theorem}

Consider a Poincar\'e invariant local QFT in a state $|\Omega \rangle$ that:
\begin{enumerate}
\item 
Does not break time translations. This means that $|\Omega \rangle$ is an eigenstate of the Hamiltonian. For simplicity, we can add a constant offset to all energies to set that of $|\Omega \rangle$ to zero.
\item
Does not break spatial translations. This means that $|\Omega \rangle$ is an eigenstate of momentum. For simplicity, we consider only cases in which the corresponding eigenvalue is zero, but it is immediate to generalize what follows to more general cases.
\item
Breaks boosts.
\end{enumerate}
Such a state could be the ground state, or the ground state at a definite chemical potential or density for some conserved charge, or a more general state with the above properties. Below, we will refer to $| \Omega \rangle$ as the `reference state'.

The fact that $| \Omega \rangle$ breaks boosts operationally means that there exists a local operator ${O}(x)$ in a non-trivial representation of Lorentz such that its variation under an infinitesimal boost has a non-zero expectation value on $| \Omega \rangle$:
\be \label{vev}
\langle \Omega | \, \delta_{ K^i} {O}(x) \,  | \Omega \rangle \equiv  i \langle \Omega | \big[ K^i, {O}(x) \big] | \Omega \rangle \neq 0 \; ,
\ee
where $K^i$ is the boost generator.
We will be more concrete about the possible choices for ${O}(x)$ below, but for now we can keep such an operator generic, including also the representation of Lorentz it belongs to. For simplicity though, we will restrict to cases where ${O}(x)$ is hermitian.

Let us work in Heisenberg picture. Then $| \Omega \rangle$ is time-independent but operators in general depend on time. However, since Lorentz is a symmetry of the dynamics, $K^i$ is conserved, that is, time-independent in Heisenberg picture.\footnote{Notice that $K^i$ does {\em not} commute with the Hamiltonian and is also {\em explicitly} time-dependent---see eq.~\eqref{Ki}. These two sources of time-dependence for $K^i$ cancel each other, thus ensuring that $K^i$ is conserved.} 
Then, since $| \Omega \rangle$ does not break space or time translations, the expectation values in \eqref{vev} must be independent of the space-time coordinates $x$:
\be \label{vevconst}
 \langle \Omega | \big[ K^i, {O}(x) \big] | \Omega \rangle \equiv -iA  \; , \qquad A = {\rm const} \in \mathbb{R} \; .
\ee
For simplicity, let us then set  $x$ to zero.

We can now use the explicit expression for the boost generators,
\be \label{Ki}
K^i = P^i t - \int d^3 x' \, x' {}^i T^{00}(\vec x \, ', t) \; ,
\ee
where $t$ is an arbitrary time, and $P^i$ is the momentum operator. Notice that, since $| \Omega \rangle$ does not break spatial translations, we can drop the $P^i t$ term when we plug \eqref{Ki} into \eqref{vevconst}.
We thus get
\be \label{vev integral}
\int d^3 x' x' {}^i \langle \Omega | \, T^{00}(\vec x \, ', t) \, {O}(0) \, | \Omega \rangle - {\rm c.c} = iA \; .
\ee

Consider now inserting an orthogonal complete set of states between the two operators on the l.h.s. Because of the properties of $| \Omega \rangle$, such states can be taken to be eigenstates of momentum and energy. In particular, we can parametrize them as $|n, \vec p \, \rangle$, where $\vec p$ is the momentum eigenvalue, and $n$ collectively denotes all other labels needed to uniquely identify an energy eigenstate. We will denote the corresponding energy eigenvalue by $E_n(\vec p \, )$. We normalize these states in such a way that the completeness relation reads
\be \label{complete}
\mathbb{1} = \int \frac{d^3 p}{(2\pi)^3} \sum_n | n, \vec p \, \rangle \langle n , \vec p \, | \; , 
\ee
where the sum over $n$ can include integrals over continuous variables such as relative momenta in multi-particle states. Notice that for single-particle states the above normalization is the so-called non-relativistic one, $\langle \vec p \, | \vec p \, ' \rangle = (2\pi)^3 \delta^3(\vec p - \vec p \, ')$. An important fact (which is relevant for other Goldstone theorems as well \cite{Nielsen:1975hm,Nicolis:2012vf}) is that our reference state $| \Omega \rangle$ cannot contribute as an intermediate state in the sum in Eq.~\eqref{vev integral}. The reason is that its contributions cancel between the two terms on the l.h.s. 

We thus have
\be 
\int \frac{d^3 p}{(2\pi)^3}   d^3 x' x' {}^i \sum_{n } \langle \Omega |  T^{00}(\vec x \, ', t) | n, \vec p \, \rangle \langle n , \vec p \, | {O}(0)  | \Omega \rangle - {\rm c.c} = iA \; ,
\ee
where we exchanged the order of integration over $\vec x \, '$ and $\vec p$.
Using the translation operators we can write
\be \label{translated T^00}
T^{00}(\vec x \, ', t) = e^{-i \vec P \cdot \vec x \,' } \,e^{+iH t} \,  T^{00}(0) \, e^{-iH t} \,e^{i \vec P \cdot \vec x \,' } \; .
\ee
Then, using properties 1 and 2 above, after straightforward algebra we get
\be \label{theorem with states}
 \lim_{\vec p \to 0} \, \frac{\partial}{\partial p ^i}   \sum_{n} e^{-iE_n (\vec p \,) t} \, {\cal T}^{00}_n(\vec p \, ) \, {\cal O}_n^*(\vec p \, ) + {\rm c.c.} = A \; ,
\ee
where the script capital letters stand for the matrix elements of the corresponding operators between $ | \Omega \rangle$ and $| n, \vec p \, \rangle $:
\be \label{T00 and O}
 {\cal T}^{00}_n(\vec p \, ) \equiv \langle \Omega | \, T^{00}(0) | n, \vec p \, \rangle \; , \qquad
 {\cal O}_n(\vec p \, ) \equiv \langle \Omega | \, O(0) | n, \vec p \, \rangle \; .
\ee
The derivative with respect to $\vec p$ is the only qualitative novelty compared to more standard Goldstone theorems, and it stems from the explicit spatial dependence of the boost charge density. Such a novelty turns out to have far reaching implications, as we will see in the following sections. 

The derivative with respect to $\vec p$  in principle can act on any of the factors inside the sum. In fact, it can also act on the {\em range} of the sum over $n$, as we will see in one of the examples below: there are systems for which the number of allowed states of total momentum $\vec p$ changes as $\vec p$ changes.
However, regardless of where it acts, the above identity implies that there are states other than $| \Omega \rangle$ whose energy goes to zero at zero momentum. The reason is that, after taking the $\vec p$ derivative and setting $\vec p$ to zero, we are left with an overall $e^{-i E_n(0) t}$ factor for each term in the sum. But the r.h.s.~of \eqref{theorem with states} is time-independent. This means that there must be states such that $E_n(0) = 0$, and that these are the only ones that contribute to the l.h.s.

Notice that so far we have not used rotational invariance in any way. {\em If} rotations are unbroken by $| \Omega \rangle$, then we can use the usual selection rules of Goldstone theorems and further conclude that the gapless intermediate states that contribute to the l.h.s.~of \eqref{theorem with states} must have zero helicity. This is because $T^{00}$ is a scalar operator under rotations, and so its matrix element in \eqref{T00 and O} is nonzero only if $|n, \vec p \, \rangle$ has zero helicity.

\subsection{Formulation in terms of spectral densities}
The  conclusion that there are gapless states is perhaps more transparent in the language of correlation functions and spectral densities. Define the Wightman correlation function
\be
G_{T^{00}, O} (x-y) \equiv \langle \Omega | T^{00} (x) \, O(y) | \Omega \rangle \; .
\ee
Then, Eq.~\eqref{vev integral} reads
\be
\int d^3 x' x' {}^i \, \big[G_{T^{00}, O} (\vec x \, ',t) - G^*_{T^{00}, O} (\vec x \, ',t)\big] = iA \; ,
\ee
which in Fourier space becomes
\be \label{theorem with G}
 \lim_{\vec p \to 0} \, \frac{\partial}{\partial p ^i} \Big[ \tilde G_{T^{00}, O} (\vec p ,\omega) - \tilde G_{T^{00}, O} ^* (-\vec p, -\omega) \Big]= A \, (2\pi) \delta(\omega)\,.
\ee
Now, it so happens that in Fourier space Wightman correlation functions are proportional to their spectral densities, with no convolution needed; we review this and the systematics of spectral densities for non-relativistic theories or states in Appendix \ref{spectral}.  More concretely, for any two local operators $O_a$, $O_b$, one simply has 
\be
\tilde G_{ab} (\vec p, \omega) = \frac{1}{2\omega} \rho_{ab} (\vec p, \omega) \; ,
\ee
with the spectral density $\rho_{ab}$ normalized in such a way that for a free, canonically normalized scalar field $\phi(x)$ that annihilates single particle states with generic dispersion relation $E = E(\vec p \,)$, the $\phi\phi$ spectral density is 
\be
\rho_{\phi\phi} (\vec p, \omega)= (2\pi) \delta(\omega - E(\vec p \,)) \; .
\ee
Then, Eq.~\eqref{theorem with G} is directly a statement about the $T^{00}$-$O$ spectral density:
\be \label{theorem with rho}
 \lim_{\vec p \to 0}  \frac{1}{2\omega} \frac{\partial}{\partial p ^i} \Big[\rho_{T^{00}, O} (\vec p ,\omega) - \rho^*_{T^{00}, O} (-\vec p, -\omega) \Big] = A \, (2\pi) \delta(\omega) \; .
\ee
Clearly, this implies that the spectral density must have support concentrated at $\omega = 0 $ for $\vec p$ going to zero, in agreement with our analysis of Eq.~\eqref{theorem with states}, that is, with the conclusion that there are gapless states. 

Again, the only qualitative difference with more standard Goldstone theorems is the derivative with respect to $\vec p$ on the l.h.s.
In those cases, the appearance of the $\delta(\omega)$ on the r.h.s.~is usually taken as an indication that the low-energy states that saturate the Goldstone theorem become stable single-particle states at low momenta. Strictly speaking, such a conclusion is justified only in  relativistic cases, that is, when $| \Omega \rangle$ is a Poincar\'e invariant vacuum of a relativistic theory. In practice however, to the best of our knowledge, that conclusion turns out to be correct also for spontaneous symmetry breaking in non-relativistic situations. For instance, for superfluid helium-4 at zero temperature, a finite-momentum phonon has a decay rate $\Gamma \sim p^5$ \cite{Lifshitz}, and so it becomes more and more stable at lower and lower momenta, in the sense that $\Gamma/\omega \to 0$ for $\vec p \to 0$.

So, the natural question is whether in our case as well we expect the identities \eqref{theorem with states} and \eqref{theorem with rho} to be saturated by zero-energy single-particle states---that is, Goldstone bosons. Although we do not have a clear understanding of why the $\vec p \,$-derivative can make such a significant difference, the answer is definitely `no'. We already saw in the previous Section that there is a tension between the usual properties of Goldstone bosons and the spontaneous breaking of boosts. In fact, in Section~\ref{sec:FL} we will see that for Fermi liquids the identities above are saturated not by Goldstone bosons---there aren't any---but rather by \emph{particle-hole} states, which form a continuum, analogous to a multi-particle continuum.
Before we do so, let us demonstrate how cases 2 and 3 discussed at the end of Sec.~\ref{heuristic}, which feature Goldstone bosons for boosts, agree with our theorem. We leave out case 1, because it involves a spontaneous breaking of spatial translations, and our theorem would need to be modified to accommodate that. 
We start with a brief digression about order parameters.

\subsection{Convenient order parameters}
So far we have been generic about the nature of our order parameter $O(x)$, but, depending on the system, there can be obviously convenient choices.

If a system features a conserved internal charge and is in a state of finite density for it, we can use the associated Noether current $J^\mu$ as an order parameter. More precisely, we can take $O(x) = J^j(x)$, so that
\be
\langle \Omega | \big[ K^i, {J^j}(x) \big] | \Omega \rangle = - i  \langle \Omega | J^0(x) | \Omega \rangle\delta^{ij} = -i \, n \, \delta^{ij}  \; ,
\ee
where $n\equiv\langle\Omega|J^0(x)|\Omega\rangle$ is the charge density of $| \Omega \rangle$.
In that case, our theorem only involves matrix elements and expectation values of conserved currents---$T^{\mu\nu}$ and $J^\mu$:
\be \label{theorem with J}
\lim_{\vec p \to 0} \, \frac{\partial}{\partial p ^i}   \sum_{n \neq \Omega} e^{-iE_n (\vec p \,) t} \, {\cal T}^{00}_n(\vec p \, ) \, {\cal J}^j_n {}^*(\vec p \, ) + {\rm c.c.} = n \, \delta^{ij} \; ,
\ee
or, in the language of spectral densities,
\be
\lim_{\vec p \to 0}  \frac{1}{2\omega} \frac{\partial}{\partial p ^i} \Big[\rho_{T^{00}, J^j} (\vec p ,\omega) - \rho^*_{T^{00}, J^j} (-\vec p, -\omega) \Big] = n \, \delta^{ij} (2\pi) \delta(\omega) \; .
\ee

However, even if our system does not feature internal charges, or if our state $| \Omega \rangle$ does not turn on any of the corresponding densities, the stress energy tensor itself is usually a good order parameter. More precisely, choosing $O(x) = T^{0j}(x)$, we have
\be\label{theorem Tmn 0}
\langle \Omega | \big[ K^i, T^{0j}(x) \big] | \Omega \rangle = - i  \langle \Omega | \,  T^{00}(x) \delta^{ij} + T^{ij}(x) \, | \Omega \rangle = - i \, (\rho + P) \, \delta^{ij}  \; ,
\ee
where $\rho$ and $P$ are the energy density and pressure of $| \Omega \rangle$, as in Eq.~\eqref{vevs}.
(Although not necessary for our theorem, here we are assuming that $| \Omega \rangle$ does not break rotations.)
So, $T^{\mu\nu}$ is not a good order parameter only for Lorentz-violating states that have vanishing $\rho+P$---such as the framid's ground state---\emph{i.e.}, only when $\langle \Omega | T^{\mu\nu} | \Omega \rangle \propto \eta^{\mu\nu}$, which is clearly exceptional for a Lorentz-violating state $| \Omega \rangle$. In all other cases, $T^{\mu\nu}$ is a particularly convenient choice of order parameter, because it exists for all local theories, and because with this choice our theorem only involves matrix elements and expectation values of $T^{\mu\nu}$ itself:
\be
\lim_{\vec p \to 0} \, \frac{\partial}{\partial p ^i}   \sum_{n \neq \Omega} e^{-iE_n (\vec p \,) t} \, {\cal T}^{00}_n(\vec p \, ) \, {\cal T}^{0j}_n {}^*(\vec p \, ) + {\rm c.c.} = (\rho+P) \, \delta^{ij} \; , \label{theorem Tmn}
\ee
or, equivalently,
\be
\lim_{\vec p \to 0}  \frac{1}{2\omega} \frac{\partial}{\partial p ^i} \Big[\rho_{T^{00}, T^{0j}} (\vec p ,\omega) - \rho^*_{T^{00}, T^{0j}} (-\vec p, -\omega) \Big] = (\rho + P) \, \delta^{ij} (2\pi) \delta(\omega) \; .
\ee

Finally, one is often interested in a non-relativistic system, by which we mean a state in a relativistic field theory that features a mass density (times $c^2$) much bigger than the typical binding energy density, kinetic energy density, pressure, etc., and propagation speeds for excitations much smaller than that of light. This is the case for typical condensed matter systems in the lab. In those cases, both  choices above are available, and are in fact equivalent. This is because in the non-relativistic limit there is automatically a new conserved charge, the total mass $M$, whose  density $J^0$ and current density $\vec J$ obey a continuity equation,
\be
\dot J^0 + \vec \nabla \cdot \vec J = 0 \; .
\ee
But $J^0$ and $\vec J$ can also be thought of as the non-relativistic limits of suitable entries of $T^{\mu\nu}$:
\be
\mbox{(NR limit)} \qquad \qquad\qquad T^{00} \simeq J^0 \; , \qquad T^{0i} \simeq J^i \; , \qquad T^{ij} \ll J^0 \; .\qquad \qquad\qquad \qquad\qquad
\ee
From the expressions above we thus get for both \eqref{theorem with J} and \eqref{theorem Tmn}
\be
\mbox{(NR limit)} \qquad 
\lim_{\vec p \to 0} \, \frac{\partial}{\partial p ^i}   \sum_{n \neq \Omega} e^{-iE_n (\vec p \,) t} \, {\cal J}^{0}_n(\vec p \, ) \, {\cal J}^{j}_n {}^*(\vec p \, ) + {\rm c.c.} = \rho_m  \, \delta^{ij} \; , \qquad \rho_m \equiv \langle \Omega | J^0(x) | \Omega \rangle\; ,
\ee
or, equivalently,
\be
\mbox{(NR limit)} \qquad \qquad  
\lim_{\vec p \to 0}  \frac{1}{2\omega} \frac{\partial}{\partial p ^i} \Big[\rho_{J^{0}, J^{j}} (\vec p ,\omega) - \rho^*_{J^{0}, J^{j}} (-\vec p, -\omega) \Big] = \rho_m \, \delta^{ij} (2\pi) \delta(\omega) \; . \qquad\qquad
\ee

\section{Systems with Goldstones}\label{sec:with goldstones}

\subsection{Framids} 
A (type-I) framid is a hypothetical phase of matter that breaks Lorentz boosts but no other symmetry, and that features Goldstone fields non-linearly realizing the broken boosts. 
The low-energy effective theory for such a system has been developed in \cite{Nicolis:2015sra}. It involves three gapless boost Goldstone fields making up a 3-vector, $\vec \eta(x)$, whose longitudinal and transverse parts  both have linear dispersion relations, generically with different speeds, and pion-like derivative interactions. Their dynamics are governed by an effective action which at quadratic  order takes the form
\be\label{action_framid}
S_\eta=\int d^4x \, \sfrac{1}{2}\left[M_1^2(\dot{\vec \eta} \, )^2-M_2^2\partial_i\eta^j\partial_i\eta^j-M_3^2(\partial_i\eta^i)^2\right]\,,
\ee
where $M_i$ are arbitrary mass scales. Finding the corresponding stress--energy tensor is more straightforward when writing the low energy action in terms of the order parameter. We shall discuss a convenient choice of order parameter below and refer the reader to the Appendix~A of \cite{Nicolis:2015sra} for the exact expressions for the stress tensor. The fields $\eta^i(x)$ admit the standard mode expansion
\be\label{eta_mode}
\eta^i(x)= \frac{1}{M_1} \int\frac{d^3p}{(2\pi)^3}\sum_h\frac{1}{\sqrt{2E_{h}(\vec p \, )}} \left[a^h_{\vec p}\,\varepsilon^i_{h,\vec p} \, e^{-ip \cdot x}+a^{h \dagger}_{\vec p}\varepsilon^{i \, *}_{h, \vec p} \, e^{ip \cdot x}\right]\,,
\ee
where $a^{\dagger h}_{\vec p},a_{\vec p}^h$ are the creation and annihilation operators, $h$ stands for the helicities $0,\pm1$, and $\varepsilon^i_{h,p}$ denote the associated (orthonormal) polarization vectors.

To see how such a system obeys our theorem through single-particle Goldstone states, we need:
\begin{enumerate}
\item An order parameter: as already emphasized, $T^{\mu\nu}$ is of no use here, because it has a Lorentz-preserving expectation value on the ground state, $\langle T^{\mu\nu}\rangle \propto \eta^{\mu\nu}$. For simplicity we will assume that the microscopic theory features a local 4-vector operator $V^\mu(x)$ with a Lorentz-breaking expectation value on $| \Omega \rangle$, 
\be
\langle \Omega | V^\mu(x) | \Omega \rangle =  \delta^{\mu}_0 \; ,
\ee
although one can generalize the analysis that follows to more complicated representations of Lorentz as well. $V_\mu$ could be a conserved Noether current, or a more general vector operator. Then, setting $O(x) = V^j(x)$ in our theorem, eq.~\eqref{theorem with states} becomes similar to \eqref{theorem with J}:
\be \label{theorem framid}
\lim_{\vec p \to 0} \, \frac{\partial}{\partial p ^i}   \sum_{n \neq \Omega} e^{-iE_n (\vec p \,) t} \, {\cal T}^{00}_n(\vec p \, ) \, {\cal V}^j_n {}^*(\vec p \, ) + {\rm c.c.} = \delta^{ij} \; ,
\ee
where the r.h.s~comes from $i \langle \Omega | \big[ K^i, {V^j}(x) \big] | \Omega \rangle =  \langle \Omega | V^0(x) | \Omega \rangle\delta^{ij} =  \delta^{ij}$.
Eq.~\eqref{theorem framid} is the identity that we need to check. In particular, we will see that it is obeyed if we restrict the sum to just single-particle Goldstone states.

\item The matrix elements of the order parameter between the ground state and single particle states: to this end, we need the expression of $V_{\mu}(x)$ to first order in the Goldstone fields obtained by performing an infinitesimal boost transformation on the expectation value of the order parameter:
\be
V^{\mu} (x)=\left(e ^{i\eta^i(x)K^i}\right)^\mu\,_\nu \langle \Omega | V^\nu(x) | \Omega \rangle \simeq \delta^{\mu}_0  + \delta^\mu_i \eta^i (x) \; .
\ee

\item The matrix elements of $T^{00}$ between the ground state and single particle states: to this end, we need the expression of $T^{00}$ to first order in the Goldstone fields. This is \cite{Nicolis:2015sra}
\be
T^{00}(x) \simeq \Lambda +  M_1^2  \, \vec \nabla \cdot \dot{\vec \eta} (x) \; , \label{T00 framid}
\ee
where $\Lambda$ is the ground state's energy, and $M_1$ is the scale appearing as an overall factor in the quadratic Lagrangian \eqref{action_framid} for $\vec \eta$. Notice that, because of the spatial divergence in \eqref{T00 framid}, only the longitudinal Goldstone states (with $h=0$) will contribute to the matrix elements we are interested in.
\end{enumerate}

Given all of the above, if we use single-particle longitudinal Goldstone states $|\vec p, L \rangle\equiv a_{\vec p,h=0}^\dagger|0\rangle$ as intermediate states for our theorem, we have:
\be
{\cal T}^{00}_{L}(\vec p \, ) = M_1 \sqrt{\frac{E_L(\vec p \,)}{2}} \, |\vec p \,|  \; , \qquad  \vec {\cal V}_L(\vec p \, ) = \frac{1}{M_1} \frac1{\sqrt{2 E_L(\vec p \,)}} \, \hat p \; , 
\ee
obtained by evaluating ${\cal T}^{00}_{L}(\vec p \, ) $ and $\vec {\mathcal V}_L(\vec p \,)$ by using the mode expansion \eqref{eta_mode}. As anticipated, we see that eq.~\eqref{theorem framid} is saturated  by these states.

\subsection{Superfluids}
From a quantum field theory standpoint, a superfluid is a system with a continuous internal symmetry in a state that on the one hand spontaneously breaks that symmetry, and on the other hand has a finite density for the associated conserved charge \cite{Son:2002zn}. This characterization is equivalent to the traditional one in terms of Bose-Einstein condensation for free bosons \cite{Lifshitz}, but, unlike that one, it has the advantage of being perfectly well defined for interacting theories as well, even at strong coupling.

One can think of the ground state of a superfluid at some chemical potential $\mu$ as the lowest-lying eigenstate of the modified Hamiltonian \cite{Nicolis:2012vf}
\be
\bar H \equiv H - \mu Q \; ,
\ee 
where $H$ is the Hamiltonian, and $Q$ the conserved charge in question. We will call such a state $| \Omega \rangle$, and use it as reference state for our Goldstone theorem. However, the requirement that $Q$ be spontaneously broken implies that $H$ also is. In other words, if $| \Omega \rangle$ is an eigenstate of $\bar H$ but not of $Q$, it cannot be an eigenstate of $H$. 
Thus, a superfluid violates assumption 1 of our theorem. 

Fortunately, there is  an easy fix.
The only places where the Hamiltonian operator is introduced in our theorem are:
\begin{enumerate}
\item
Implicitly, in the sum over the complete set of states of Eq.~\eqref{complete}: those states are supposed to be eigenstates of $H$. In our case, since $| \Omega \rangle$ breaks $H$ but not $\bar H$, we can classify its excitations in terms of eigenstates of $\bar H$. And, so, we can still use  Eq.~\eqref{complete}, but with the understanding that $|n, \vec p \, \rangle$ is now an eigenstate of $\vec P$ and $\bar H$, but not of $H$ (or $Q$). We will call $E_n (\vec p \,)$ the corresponding eigenvalue of $\bar H$, and refer to it as the `energy' of  $|n, \vec p \,\rangle$.
\item
Explicitly, in Eq.~\eqref{translated T^00}. However, since $T^{00}$ and $H$ are both neutral under $Q$ (\emph{i.e.}, they commute with it, because $Q$ generates an internal symmetry), that equation is equally valid if one uses $\bar H$ in place of $H$ in it:
\be
e^{+iH t} \,  T^{00}(0) \, e^{-iH t} = e^{+i \bar H t} \,  T^{00}(0) \, e^{-i \bar H t}
\ee
\end{enumerate}
We thus reach the conclusion that our theorem should be valid for superfluids as well, as long as one interprets the states and energies appearing there as eigenstates and eigenvalues of $\bar H$ rather than of $H$.\footnote{Notice that, however unsurprising it might sound, this conclusion was not obvious \emph{a priori}: our theorem has to do with boosts, and $H$ and $\bar H$ behave very differently under boosts. The former is the timelike component of a four-vector $P^\mu$, the latter is a linear combination of this and of a scalar $Q$.}

With these qualifications in mind, we can now check that, for superfluids, single-particle phonon states saturate our Goldstone 
theorem for boosts. The effective field theory for low-energy excitations of relativistic superfluids was developed by Son \cite{Son:2002zn}. It involves a Lorentz scalar $\phi(x)$ with a shift symmetry, $\phi \to \phi + {\rm const}$, expanded about the background configuration $\langle \Omega | \phi(x) | \Omega \rangle = \mu t$:
\be
\phi(x) = \mu t + \pi(x) \; ,
\ee
where $\pi(x)$ is the field that creates and annihilates phonons.

The low-energy effective action is
\be
S = \int d^4 x \, P \big(  X \big) \; , \qquad X \equiv (\di \phi)^2 \; ,
\ee
where $P(\mu^2)$ is the superfluid's equation of state, given by the pressure as a function of the chemical potential (squared).
From this, one can compute
the stress-energy tensor and the $U(1)$ current:
\be \label{T and J}
T_{\mu\nu}= 2 P'(X) \di_\mu \phi \, \di_\nu \phi - \eta_{\mu\nu} P(X) \; , \qquad J_\mu = 2 P'(X) \di_\mu \phi \; .
\ee
We can use $J^i$ as our order parameter $O$. Our theorem thus takes the form \eqref{theorem with J}. Notice that the background charge density is
\be \label{n}
n = \langle \Omega | J^0 | \Omega \rangle = 2 P'(\mu^2) \mu = \frac{d P}{d \mu} \; ,
\ee
as befits a zero-temperature superfluid.

Similarly to the framids' case, we will need the expansion of $T^{00}$ and of $J^i$ to first-order in the Goldstone field $\pi(x)$. We have:
\be
T^{00} \simeq \frac{n}{c_s^2} \dot \pi \; , \qquad J^i = - \frac{n}{\mu} \di_i \pi \; ,
\ee
where we used Eq.~\eqref{n} as well the expression for the speed of sound:\footnote{The expression for the energy density $\rho$ in terms of $\mu^2$ can be found from Eq.~\eqref{T and J} above as $\rho = T^0_0$, or, equivalently, from the zero-temperature thermodynamic identity $\rho+ P = \mu n$.}
\be
c_s^2 = \frac{d P}{ d \rho} = \frac{P'}{2 P'' \mu^2 + P'} \; .
\ee
Finally, we need to know the normalization of the $\pi$ field compared to canonical normalization. For this, we need the expansion of the effective action to quadratic order in $\pi$: 
\be
S \simeq \frac{n}{\mu c_s^2} \int d^4 x \, \frac12 \Big[ \dot \pi^2- c_s^2 \big (\vec \nabla \pi \big)^2 \Big] \; .
\ee

With this normalization of $\pi(x)$, for a single-particle phonon state $|\vec p \, \rangle $ we have
\be
\langle  \Omega | \, \pi(0) \, |\vec p \, \rangle = c_s \sqrt{\frac{\mu}{n}} \frac1{\sqrt{2 E(\vec p \,)}}  \;, \qquad E(\vec p \,) = c_s p \; , 
\ee
and thus the matrix elements relevant for our theorem  are
\be
{\cal T}^{00}(\vec p \, ) = -i \sqrt{\frac{n \mu}{c_s}} \, \sqrt{\frac{p}{2}}   \; , \qquad  \vec {\cal J}(\vec p \, ) =  -i \sqrt{\frac{n c_s}{\mu }} \, \frac{\vec p}{\sqrt{2  p}}   \; , 
\ee
in perfect agreement with Eq.~\eqref{theorem with J}.

Notice that here we have ${\cal T}^{00} \sim p^{1/2}$ and ${\cal O} = \vec {\cal J} \sim p^{1/2}$, while for the framid we had ${\cal T}^{00} \sim p^{3/2}$ and ${\cal O}  = \vec {\cal V} \sim p^{-1/2}$. In both cases ${\cal T}^{00} {\cal O}^*$ scales as $p$, as it should in order to satisfy our theorem. Still, this linear scaling with $p$ of the product is achieved in substantially  different ways in the two cases. We expect this to be related to the presence of inverse Higgs constraints for the superfluid, $\vec \eta \sim \vec \nabla \pi$, although the precise relationship is not immediately obvious to us.

\section{Systems without Goldstones: the massive particle}\label{sec:massive_part}
As a warmup for the more physically relevant example of Fermi liquids, let's consider a relativistic QFT whose lightest states are spinless massive particles of mass $m$. The theory can have generic interactions, not necessarily perturbative, as long as there exist asymptotic states, the lightest of which have mass $m$.

Now, if we  take one such particle in a state of definite momentum $\vec p$, such a state has all the properties spelled out at the outset of sect.~\ref{theorem}, and thus qualifies as a perfectly good reference state for our theorem. In fact, since the particle is massive, we can go to its rest frame, and consider the zero-momentum reference state\footnote{The volume factor upfront makes $| \Omega \rangle$ normalized to one, as implicitly assumed in our theorem. Similarly, all the other volume factors introduced in this section make  normalizations  consistent with our previous choices.}
\be
| \Omega \rangle \equiv \frac{1}{\sqrt V} | \vec p = 0 \rangle=\frac{1}{\sqrt{V}}a^\dagger_0|0\rangle \; ,
\ee
which simplifies the analysis somewhat.

How is our theorem obeyed? Clearly the theory has no gapless Goldstone bosons: the lightest single-particle states are our massive particles. However, starting from the $| \Omega \rangle$ above, we can find states that are arbitrarily close to it in energy when their momentum approaches zero: such states describe the same one particle that is already there, but with momentum slightly different from zero. In other words, they are slightly boosted versions of our reference state $| \Omega \rangle$. These are simply
\be
| \vec p \, \rangle \; , \qquad \vec p \neq 0 \; ,
\ee
and their energy with respect to $| \Omega \rangle$ in the low-momentum limit is
\be \label{E(p) massive particle}
E(\vec p \,) \simeq \frac{\vec p\, ^2}{2 m} \; , 
\ee
which goes to zero for $\vec p$ going to zero.

One could argue that these are in fact single-particle  states. They are, but only as far as  the true vacuum of the theory in concerned. Instead, with respect to our boost-breaking reference state, they are {\em particle-hole} states: to get them from $| \Omega \rangle$, we must first annihilate a particle with zero momentum, and then create one with momentum $\vec p \,$:
\be \label{states massive particle}
| \vec p \, \rangle  = \frac{1}{\sqrt V} a^\dagger_{\vec p} \, a_0 | \Omega \rangle \; .
\ee

These states do not form a continuum like more standard multi-particle or particle-hole states: for each total momentum $\vec p$, there is only one state in this class, with energy (w.r.t.~$| \Omega \rangle$) given by \eqref{E(p) massive particle}. This again may lead one to believe that these are standard single-particle states also as far as $| \Omega \rangle$ is concerned. But they are not. In fact, taking the energy of $| \Omega \rangle$ as a reference point, the spectrum of the theory is quite peculiar: we can only have {\em one} quantum whose energy goes to zero with $\vec p$, but no more. We cannot add more quanta of this excitation. Once we apply $a^\dagger_{\vec p} \, a_0$ to $| \Omega \rangle$, the only way to get another gapless excitation at small momentum is to apply something like $a^\dagger_{\vec q} \, a_{\vec p}$, which gives us again one of our states, this time with momentum $\vec q$, but that with momentum $\vec p$ is gone. All the other states correspond to adding more particles of the original theory, and thus have gap $m$. Put another way: taking $| \Omega \rangle$ as reference state, we have gapless single-quantum states, but the multi-quantum continuum is gapped.

Let's check that if we use the states \eqref{states massive particle} as the intermediate $|n,\vec p  \,\rangle$ states in our Goldstone theorem, the theorem \eqref{theorem Tmn} is obeyed. First, notice that in this case we do not need the additional label $n$ to characterize the intermediate states. Second, notice that the states are already normalized correctly. Then, we just need to compute the matrix elements
\be
{\cal T}^{\mu\nu}(\vec p \,) \equiv \langle \Omega | T^{\mu\nu}(0) | \vec p \, \rangle
\ee
for small $\vec p$. In particular, we need the terms of first order in $\vec p$ in the product ${\cal T}^{00}(\vec p \,) {\cal T}^{0j} {}^*(\vec p)$ (see eq.~\eqref{theorem Tmn}). 

In a relativistic field theory, because of Lorentz invariance and basic symmetry properties, the matrix elements of $T^{\mu\nu}$ between single-particle states (of identical spinless particles) must take the form
\begin{align}
\langle \vec k \, | T^{\mu\nu}(0) | \vec p \, \rangle = & \;  \sfrac12( k^\mu k^\nu + p^\mu p^\nu) F(k\cdot p) +
\sfrac12( k^\mu p^\nu + p^\mu k^\nu) G(k\cdot p)  \\
& + \eta^{\mu\nu} H(k\cdot p) + i \sfrac12( k^\mu k^\nu - p^\mu p^\nu) I(k\cdot p) \; ,
\end{align}
for some real functions $F$, $G$, $H$, $I$, 
and must obey the limit
\footnote{This follows from Lorentz invariance and from imposing, with our normalizations, $\int d^3 x \, \langle \vec k \, |   T^{0\mu}(x) | \vec p \,\rangle \to V \cdot k^\mu$  for $\vec p \to \vec k$. Cf.~\cite{Weinberg:1980kq}.}
\be\label{Tmn_single_part}
\langle \vec k \, | T^{\mu\nu}(0) | \vec p \, \rangle \to \frac{k^\mu k^\nu}{k^0} \quad {\rm for} \: \vec p \to \vec k \; .
\ee
This implies 
\be
F(m^2) + G(m^2) = \frac 1 {k^0} \; , \qquad H(m^2) = 0 \; .
\ee
Furthermore, conservation of $T^{\mu\nu}$, in the form $[P_\mu, T^{\mu\nu}] = 0$, implies
\be
\langle \vec k \, | T^{\mu\nu}(0) | \vec p \, \rangle \, (k_\mu - p_\mu) = 0 \; , 
\ee
that is, 
\be
I(k\cdot p) = 0 \; , \qquad H(k \cdot p)  = \big(F(k \cdot p) - G(k \cdot p) \big) \cdot (k\cdot p-m^2) \; .
\ee

For us, to the order we are interested in, all this implies
\be
{\cal T}^{00}(\vec p \,) \simeq  \frac{1}{\sqrt V} m \; , \qquad {\cal T}^{0j}(\vec p \,) \simeq  \frac{1}{\sqrt V} \frac12 \, p^j \; ,
\ee
which, plugged into the l.h.s.~of eq.~\eqref{theorem Tmn}, yields
\be
 \lim_{\vec p \to 0} \, \frac{\partial}{\partial p ^i}  \big[  e^{-iE (\vec p \,) t} \, {\cal T}^{00}(\vec p \, ) \, {\cal T}^{0j} {}^*(\vec p \, ) + {\rm c.c.}\Big] = \frac{m}{V} \delta^{ij} \; .
\ee
This is precisely the expected result according to eq.~\eqref{theorem Tmn}: since we have a single particle at rest in a volume $V$, the expectation value of $T^{00}$ is $m/V$, and that of $T^{ij}$ vanishes.

This is admittedly a fairly degenerate example: in the infinite volume limit, all the physical effects of having a single zero-momentum particle rather than the vacuum must be infinitely diluted. For instance, assuming the theory has local quartic interactions of some kind (e.g.~$\lambda \phi^4$), the two-to-two cross section for particles propagating in the true vacuum of the theory is finite,
\be
d \sigma^0_{1,2 \to 3,4} = \frac{1}{2 E_1} \frac{1}{2 E_2} \frac{1}{v_{12}} |{\cal M}^0_{1,2 \to 3,4}|^2 d \Pi_{3,4} \; ,
\ee
where we are using standard relativistic scattering theory notation, and the superscript zero reminds us that the corresponding quantities are to be computed in the vacuum.
However, if one of the initial particles, say 1, is at rest, and one of the final ones, say 3, has very small momentum compared to $m$, we can think of this same process as a one-to-two process taking place in $| \Omega \rangle$, whereby gapped excitation 2 decays into gapless excitation 3 and gapped excitation 4. The rate for this one-to-two process takes the form
\be
d \Gamma^{\Omega}_{2 \to 3,4} = \frac{1}{2 E_2} |{\cal M}^\Omega_{2 \to 3,4}|^2 d \Pi_{3,4} \; ,
\ee
where the superscript $\Omega$ reminds us that, now, the corresponding quantities are to be computed in $| \Omega \rangle$. In particular, $|2\rangle\equiv a^\dagger_2|\Omega\rangle=\frac{1}{\sqrt{V}}a^\dagger_2a^\dagger _0|0\rangle$ and $\langle 3, 4|\equiv \langle \Omega|\frac{1}{\sqrt{V}}a^\dagger_0a_3a_4=\langle 0|a_3 a_4$. And this is where we see the dilution phenomenon alluded to above: the physical process is exactly the same, but the kinematical factors we have to use for correctly normalizing the amplitude ${\cal M}$ are different in the two cases---they depend on the number of external legs. Following standard scattering theory, for our kinematics we have 
\be
{\cal M}^\Omega_{2 \to 3,4} = \frac{1}{\sqrt{2 m \, V}} {\cal M}^0_{1,2 \to 3,4} \; ,
\ee
and so
\be
d \Gamma^{\Omega}_{2 \to 3,4} = \frac{1}{V} \times v_2 \, d \sigma^0_{1,2 \to 3,4} \; ,
\ee
which goes to zero at infinite volume, as expected.

Yet, this perhaps academic example serves to illustrate that systems that spontaneously break boosts and no other symmetries can have quite peculiar spectra of excitations.

\section{Fermi liquids}\label{sec:FL}
Fermi liquid theory is normally used to describe electrons in metals at low temperatures. There, however, boost invariance is broken by the underlying solid, whose phonons can serve as Goldstone bosons both for translations and for boosts. Here instead, we are interested in cases in which it is the ground state of the Fermi liquid itself that breaks boosts while preserving translational invariance. The prime example of this is liquid helium-3, at temperatures low enough so that Fermi degeneracy is more important than thermal fluctuations ($T \lesssim 1$ K), but high enough so that Cooper-pairing and the associated onset of superfluidity can be neglected ($T \gtrsim 1$ mK).
Fermi liquid theory assumes that the ground state of a Fermi liquid consists of fermionic particles that occupy all momentum states with $|\vec p \, |\leq p_{\rm F}$, as is the case also for a free Fermi gas. It then further assumes that these \emph{quasiparticles} in an energy band close to the Fermi surface (\emph{i.e.} $|\vec p \, |=p_{\rm F}$) are  weakly interacting (this is shown in Fig.~\ref{Fig:fermi} [left]). This theory correctly predicts the spectrum of lowest-lying excitations in Fermi liquids, consisting of particle-hole continuum and  first and zero sound modes \cite{Lifshitz,Abrikosov}.  From an effective field theory viewpoint, Landau's theory can be reinterpreted as a theory of almost free fermionic particles with momentum close to the Fermi momentum $p_{\rm F}$ with an irrelevant quartic interaction that only becomes marginal for specific momentum configurations \cite{Polchinski:1992ed}. Despite their apparent simplicity, there is no local position space Lagrangian description of Fermi liquids and their low-energy dynamics. As we shall see below, the Goldstone theorem \eqref{theorem Tmn} presented in this work can in turn be equally well used also for Fermi liquids even in the absence of a local field theory description.

Importantly in the context of this paper, the ground state of a Fermi liquid is a state of finite energy density and low pressure, thus breaking  boosts.
We define a Fermi liquid's ground state as a tensor product of single-particle momentum eigenstates: 
\be \label{FL state}
|\text{FL}\rangle\equiv \mathcal N \prod_s\prod_{|\vec p \, |\leq p_{\rm F}}|\vec p\,\rangle\,,\qquad |\vec p\,\rangle\equiv c^s_{\vec p}\,^\dagger|0\rangle\,,
\ee
where $c^s_{\vec p}\,^\dagger$ is the fermionic creation operator creating a single particle state of momentum $\vec p$ and spin $s$, satisfying the anticommutation relation $\{c^s_{\vec p},c^{s'\dagger}_{\vec p \, '}\}=(2\pi)^3\delta^{(3)}(\vec p - \vec p \, ')\delta_{ss'}$. We emphasize that the $c^\dagger$'s and $c$'s create and annihilate quasiparticle states, which are dressed versions of the single-particle states defined on the vacuum state $| 0 \rangle$ of the microscopic theory. Such `dressing' depends both on the interactions of the microscopic theory and, crucially, on the density of particles present in the state we are considering, which is $| {\rm FL} \rangle$ itself. So, \eqref{FL state} is an implicit definition of our state. Landau's Fermi liquid theory essentially assumes that such a definition in terms of quasiparticle states is possible (see, \emph{e.g.}~\cite{Varma}). 
The normalization factor $\mathcal N$ is chosen so that $\langle\text{FL}|\text{FL}\rangle=1$. Due to the anticommutation properties, it is clear that there can in fact only be one-particle states inside our Fermi liquid ground state, which further satisfies 
\begin{align}\label{vanishFL}
c^s_{\vec p}\,^\dagger|\text{FL}\rangle&=0\,,\qquad |\vec p \, |\leq p_{\rm F}\,,\\
c^s_{\vec p}\,|\text{FL}\rangle&=0\,,\qquad |\vec p \, |> p_{\rm F}\,.
\end{align}
The first equality says that one cannot create another copy of a particle already contained in the Fermi liquid ground state:  this is the Pauli exclusion principle, and follows trivially from the anticommutation relation between the fermionic creation and annihilation operators.
The second equality instead says that one cannot annihilate a particle that is not contained in the Fermi liquid ground state already. 

\begin{figure}[t]
    \centering
\includegraphics[width=6cm]{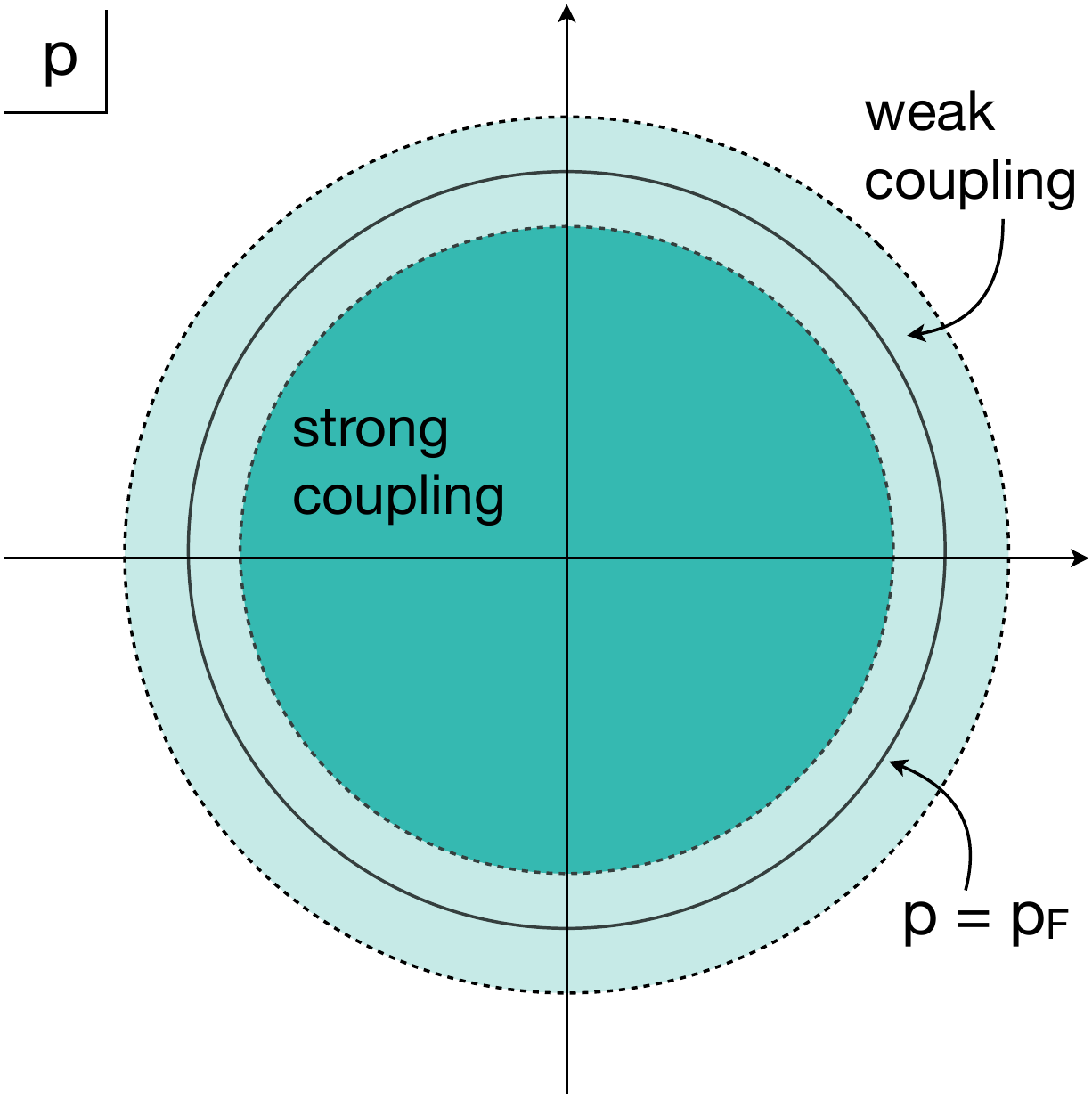}
\qquad
\includegraphics[width=6cm]{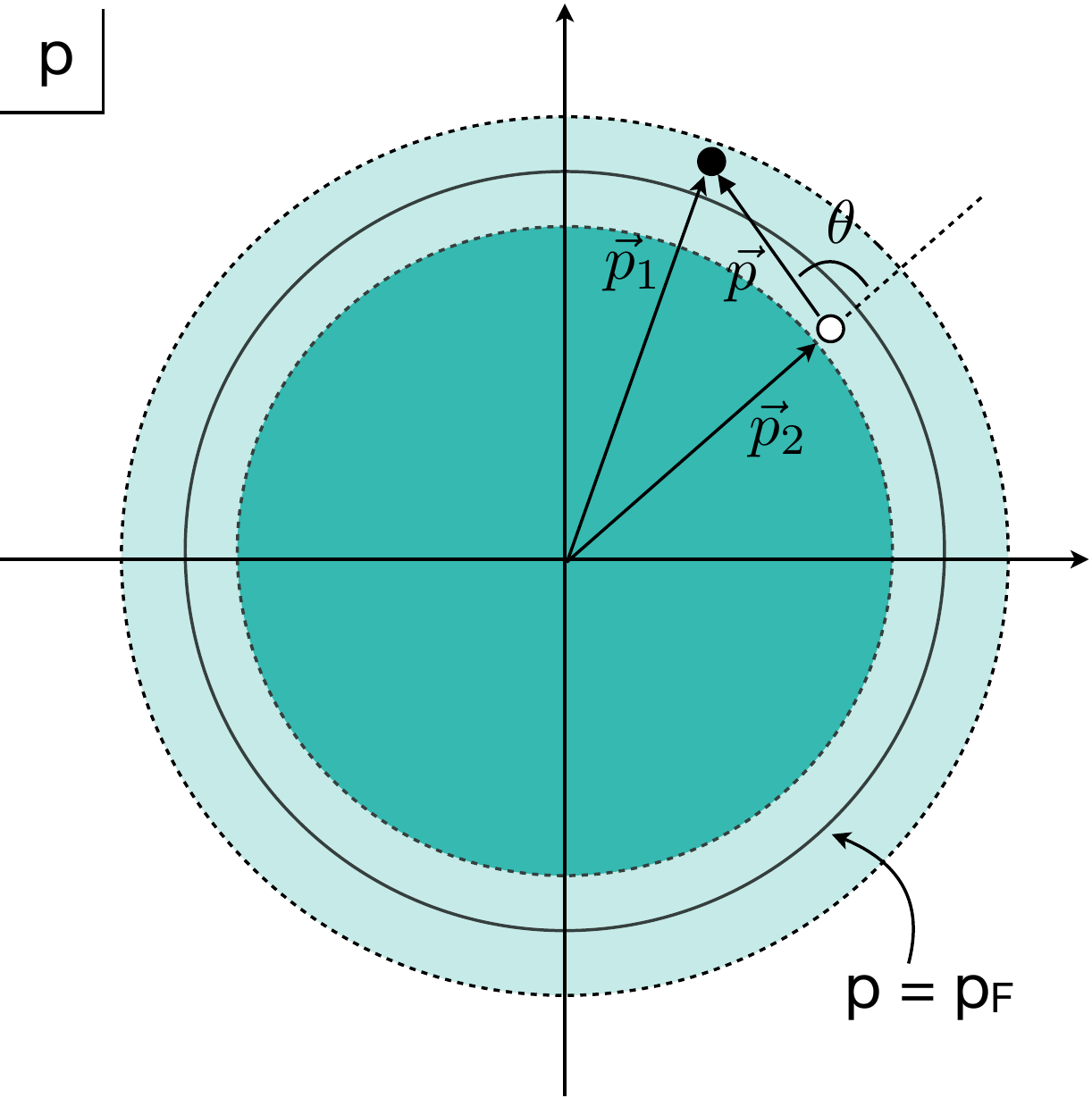}
\caption{The momentum space Fermi surface. \textbf{Left:} The weak (light green) and strong (dark green) coupling regions. \textbf{Right:} The particle--hole state consisting of a hole (empty dot) with $\vec p_2$ and a particle (full dot) with $\vec p_1$.}
\label{Fig:fermi}
\end{figure}

One can define a \emph{particle} as a one-particle state on top of the Fermi liquid:
\be
|\mathbb 1_{\vec p}^s \rangle\equiv c^s_{\vec p}\,^\dagger|\text{FL}\rangle\,,\qquad |\vec p \, |> p_{\rm F}\,.
\ee
Similarly, a \emph{hole} is defined as a Fermi liquid missing one particle of given momentum:
\be
|\bar{\mathbb 1}_{\vec q}^s \rangle\equiv c^s_{\vec q}\,|\text{FL}\rangle\,,\qquad |\vec q \, |\leq p_{\rm F}\,.
\ee
We can now introduce the concept of a \emph{particle-hole} state as
\be\label{PH_state}
|\psi\,\rangle\equiv {c^{s\,\dagger}_{\vec p_1}}\,c^{s'}_{\vec p_2}|\text{FL}\rangle\,,
\ee
where the annihilation operator creates a hole with a momentum $\vec p_2, \,|\vec p_2|\leq p_{\rm F}$ and spin $s'$ and the creation operator creates a particle with a momentum $\vec p_1,\,|\vec p_1|>p_{\rm F}$ and spin $s$. The total momentum characterising the state $|\psi\,\rangle$ is then given by the difference $\vec p\equiv \vec p_1-\vec p_2$. This configuration is shown in Fig.~\ref{Fig:fermi}. 
Although the total momentum of a low-energy particle-hole state can be as large as $|\vec p \, | \simeq 2p_{\rm F}$, for the purpose of the Goldstone theorem eventually we will be interested in the $|\vec p \, |\ll p_{\rm F}$ case. 
In such a limit, the  energy of a particle-hole state is $E(\vec p \,,\vec p_2)\equiv E(p_1)-E(p_2) \simeq \partial_{\vec p_2}E(\vec p_2)\vec p\equiv v_{\rm F}\,\hat p_2\cdot\vec p$, where $\hat p_2$ is the unit vector in the direction of $\vec p_2$ and $v_{\rm F}$ is the so-called Fermi velocity \cite{Polchinski:1992ed}. 

We shall argue below that, in the case of Fermi liquids, our boost Goldstone theorem is obeyed by the particle-hole states \eqref{PH_state}. 
Restricted to them, the completeness relations reads
\be \label{complete2}
\mathbb{1} \supset \sum_{s,s'}\int \frac{d^3 p_1}{(2\pi)^3} \,\frac{d^3p_2}{(2\pi)^3} | \psi \, \rangle \langle \psi \, |=\sum_{s,s'}\int \frac{d^3 p}{(2\pi)^3} \,\frac{d^3p_2}{(2\pi)^3} | \psi \, \rangle \langle \psi \, | \; , 
\ee
where $p$ in the last equality is the total momentum of the state $|\psi\rangle$ and we have chosen to leave $p_2$ as the other integration variable. The theorem \eqref{theorem Tmn} 
then takes the form\footnote{We omit the sum over spins for simplicity here, but will restore it in our explicit computations below where necessary. In general, the spin indices always appear hand in hand with the associated momenta and can thus be restored at the end of computations without any ambiguity.}
\be
\lim_{\vec p \to 0} \, \frac{\partial}{\partial p ^i}   \int \frac{d^3p_2}{(2\pi)^3}\, e^{-iE(\vec p,\vec p_2) t} \, {\cal T}^{00}(\vec p \,,\vec p_2 ) \, {\cal T}^{0j} {}^*(\vec p \,,\vec p_2 ) + {\rm c.c.} = (\rho_{\rm FL}+P_{\rm FL}) \, \delta^{ij} \; , \label{theorem Tmn FL}
\ee
where now the matrix elements are defined as
\be\label{matrix_elem}
{\cal T}^{\mu\nu}(\vec p \,,\vec p_2) \equiv \langle \text{FL} | T^{\mu\nu}(0) | \psi \, \rangle\,.
\ee

The technical complication when discussing Fermi liquids is that there is no known local position space quantum field theory describing them. In particular, 
there is no local expression for the stress-energy tensor operator in terms of the fields describing the low-energy excitations of the system. 
However, we do know that in the ground state the stress-energy tensor satisfies
\be\label{fermi_gs}
\langle \text{FL}|T^{00} |\text{FL}\rangle = \rho_{\rm{FL}} \; , \qquad\langle \text{FL}|T^{0i} |\text{FL}\rangle = 0 \; , \qquad  \langle \text{FL}|T^{ij} |\text{FL}\rangle = P_{\rm{FL}} \,  \delta^{ij} \; ,
\ee
where $\rho_{\rm{FL}}$ and $P_{\rm{FL}}$ are the ground state energy density and pressure. In the non-relativistic limit one has $P_{\rm{FL}}\ll\rho_{\rm{FL}}$. Furthermore, according to Landau's theory, in a momentum band close to the Fermi surface (see Fig.~\ref{Fig:fermi}), Fermi liquid is well described by weakly interacting quasiparticles.  We therefore split the stress-energy tensor describing a Fermi liquid as 
\be\label{Tmn split}
T^{\mu\nu}(x)=T^{\mu\nu}_{\text{free}}(x)+T^{\mu\nu}_{\text{interacting}}(x)\,.
\ee
In the following subsections we focus on the structure of the free and interacting parts of the stress--energy tensor. As we shall see, there is no actual need for separating it into a free part and an interaction part. Nevertheless, it is instructive to do so in order to distinguish between the properties of bosonic and fermionic quasiparticles. For the benefit of the impatient reader, we provide in Appendix \ref{quick} a quick check of our theorem for Fermi liquids that does not require inspecting the structure of the stress-energy tensor too closely.

\subsection{Free theory}
\subsubsection{Brute force}
Given that Fermi liquid theory is an effective theory usually describing systems of strongly interacting fermionic particles (\emph{e.g.} electrons in metals), it is natural to assume that the corresponding quasi-particles are fermionic. This is, in fact, what has been done in \cite{Polchinski:1992ed} where  Landau's theory was rederived from an EFT perspective. Hence using the stress-energy tensor of free fermions as the $T^{\mu\nu}_{\text{free}}(x)$ in \eqref{Tmn split} is certainly plausible. On the other hand, some quantum liquids, \emph{e.g.}~liquid helium-4, can be described as interacting Bose gases \cite{Lifshitz}. We shall therefore consider both bosonic and fermionic quasiparticles in what follows.

\subsubsection*{Bosons}\label{free bosons}
One can in fact define a Fermi liquid-like state by using bosonic creation and annihilation operators:
\be\label{FL_boson}
|\text{FL}\rangle\equiv \mathcal N \prod_{|\vec p \, |\leq p_{\rm F}}|\vec p\,\rangle\,,\qquad |\vec p\,\rangle\equiv a_{\vec p}^\dagger|0\rangle\,,
\ee
where similarly as before the creation and annihilation operators satisfy the commutation relation $[a_{\vec p},a^\dagger_{\vec p\,'}]=(2\pi)^3\delta^{(3)}(\vec p - \vec p\,')$ and the normalization factor $\mathcal N$ is chosen so that $\langle\text{FL}|\text{FL}\rangle=1$. Such a state is certainly not the ground state of the theory, even at finite chemical potential. Still, for {\em free} bosons, it is an eigenstate of the Hamiltonian, and can thus be used as the reference state $|\Omega \rangle$ in our theorem.

In distinction from a Fermi liquid state built out of fermionic operators, the only relation that is satisfied by the bosonic $|\text{FL}\rangle$ is
\begin{align}\label{vanishing}
a_{\vec p}\,|\text{FL}\rangle&=0\,,\qquad |\vec p \, |> p_{\rm F}\,.
\end{align}
Acting on the Fermi liquid state with a creation operator of momentum $|\vec q \, |$ less than $p_{\rm F}$ would in turn create a state, very similar to \eqref{FL_boson}, but with one of the single particle states replaced by a two-particle state of momentum $\vec q$. As before, one can define particle and hole states as
\begin{align}
|\mathbb 1_{\vec p} \rangle\equiv a_{\vec p}^\dagger|\text{FL}\rangle\,,\qquad |\vec p \, |> p_{\rm F}\,,\\
|\bar{\mathbb 1}_{\vec q} \rangle\equiv a_{\vec q}\,|\text{FL}\rangle\,,\qquad |\vec q \, |\leq p_{\rm F}\,.
\end{align}
For later convenience, let us emphasize that the hole (and also particle) states form an orthogonal set of states, \emph{i.e.}~$\langle\text{FL}|a^\dagger_{\vec k}a_{\vec q}|\text{FL}\rangle=\tilde{\mathcal N}(2\pi)^3\delta^{(3)}(\vec k-\vec q \,)$ for $|\vec k|\,,|\vec q \, | \le p_{\rm F}$. The normalization constant can be determined by first noting that for any single particle state $|\vec p \, \rangle\equiv a^\dagger_{\vec p}|0\rangle$ it holds that $a^\dagger_{\vec q}\,a_{\vec q}|\vec p \, \rangle=(2\pi)^3\delta^{(3)}(\vec q-\vec p \, )|\vec q \, \rangle$ and in particular $a^\dagger_{\vec q}\,a_{\vec q}|\vec q \, \rangle=(2\pi)^3\delta^{(3)}(0)|\vec q \, \rangle\equiv V|\vec q \, \rangle$. Now, applying the same steps to the Fermi liquid ground state allows us to determine $\tilde{\mathcal N}=1$ and gives
\be\label{elements}
\langle\text{FL}|a^\dagger_{\vec q}\,a_{\vec q}|\text{FL}\rangle=V\,,\qquad \langle\text{FL}|a^\dagger_{\vec k}a_{\vec q}|\text{FL}\rangle=(2\pi)^3\delta^{(3)}(\vec k-\vec q)\,,\qquad |\vec k|\,,|\vec q \, |\leq p_{\rm F}\,.
\ee
Finally, the bosonic particle-hole state is defined as
\be\label{PH_state2}
|\psi\,\rangle\equiv {a^{\dagger}_{\vec p_1}}\,a_{\vec p_2}|\text{FL}\rangle\,,\qquad |\vec p_2|\leq p_{\rm F}\,,|\vec p_1|>p_{\rm F}\,.
\ee

We then take a free relativistic scalar field described by the standard action\footnote{ We remind the reader that we are using the $(+,-,-,-)$ signature. }
\be
S = \int d^4 x\sqrt{-g}\,\left[\frac{1}{2}(\partial\phi)^2-\frac{1}{2}m^2\phi^2\right]\,
\ee
giving for the Minkowski stress-energy tensor
\be
T^{\mu\nu}(x)=\partial^\mu\phi\,\partial^\nu\phi-\eta^{\mu\nu}\left(\frac{1}{2}(\partial\phi)^2-\frac{1}{2}m^2\phi^2\right)\,.
\ee
Performing the mode expansion
\be
\phi(x)=\int\frac{d^3k}{(2\pi)^3}\frac{1}{\sqrt{2\omega_k}}\left(a_{\vec k}e^{-ikx}+a^\dagger_{\vec k}e^{ikx}\right)
\ee
and evaluating the stress tensor at $x=0$ for the relevant stress tensor components gives
\be
\begin{split}
T^{00}(0)&\supset\int\frac{d^3k\,d^3q}{(2\pi)^6}\frac{1}{4}\frac{1}{\sqrt{\omega_k\omega_q}}\left(a^\dagger_{\vec k}a_{\vec q}+a_{\vec k}a^\dagger_{\vec q}\right)\left(\omega_q\omega_k+\vec k\cdot\vec q+m^2\right)\,,\\
T^{0i}(0)&\supset\int \frac{d^3k\,d^3q}{(2\pi)^6}\frac{1}{2}\sqrt{\frac{\omega_k}{\omega_q}}q^i\left(a^\dagger_{\vec k}a_{\vec q}+a_{\vec k}a^\dagger_{\vec q}\right)\,,
\end{split}
\ee
where $\supset$ indicates that we only look at the elements with equal number of creation and annihilation operators. After normal ordering,\footnote{ Here we perform the normal ordering with respect to the true vacuum as $:\mathcal O:\,\equiv\mathcal O-\langle 0|\mathcal O|0\rangle$. Alternatively one could normally order with respect to $|\rm FL\rangle$. However, the Goldstone theorem \eqref{theorem Tmn 0} is insensitive to these specifics. The reason is that, due to Lorentz invariance, the contribution to the stress-energy tensor being subtracted by normal ordering can only be proportional to $\eta_{\mu\nu}$. In other words, it can only shift the cosmological constant, for which $\rho +p=0$. Thus, it does not contribute to the right hand side of \eqref{theorem Tmn 0}. As for the left hand side, such contribution is proportional to the identity operator and  cancels out of the commutator.} we obtain in the non-relativistic limit 
\be
\begin{split}
:T^{00}_{\text{N.R.}}(0):&=m\int\frac{d^3k\,d^3q}{(2\pi)^6}a^\dagger_{\vec k}a_{\vec q}\,,\\
:T^{0i}_{\text{N.R.}}(0):&=\frac{1}{2}\int \frac{d^3k\,d^3q}{(2\pi)^6}a^\dagger_{\vec k}a_{\vec q}\,(q^i+k^i)\,.
\end{split}
\ee
In order to evaluate the matrix elements \eqref{matrix_elem} needed for the Goldstone theorem we need
\be
\begin{split}
\langle\text{FL}|a^\dagger_{\vec k} a_{\vec q}|\psi\rangle=&\langle\text{FL}|a^\dagger_{\vec k}a^\dagger_{\vec p_1}a_{\vec q}a_{\vec p_2}|\text{FL}\rangle+(2\pi)^3\delta^{(3)}(\vec q-\vec p_1)\langle\text{FL}|a^\dagger_{\vec k}a_{\vec p_2}|\text{FL}\rangle\\
=&(2\pi)^6\delta^{(3)}(\vec q-\vec p_1)\,\delta^{(3)}(\vec k-\vec p_2)\,,
\end{split}
\ee
where we have expressed the particle--hole state as in \eqref{PH_state2}. On the second equality we have used the property \eqref{vanishing} for $\vec p_1$ and the expression for matrix elements \eqref{elements}. We thus obtain
\begin{align}\label{elements}
{\cal T}^{00}(\vec p \,,\vec p_2 ) =m\,,\qquad {\cal T}^{0i}(\vec p \,,\vec p_2 ) =\frac{1}{2}(p_1^i+p_2^i)\,.
\end{align}

For the Goldstone theorem in \eqref{theorem Tmn FL} above this means that we need to evaluate
\be\label{integral}
 \int \frac{d^3p_2}{(2\pi)^3}\, {\cal T}^{00}(\vec p \,,\vec p_2 ) \, {\cal T}^{0j} {}^*(\vec p \,,\vec p_2 ) + {\rm c.c.} = m\int\frac{d^3p_2}{(2\pi)^3}\left(p_1^i+p_2^i\right)=m\int\frac{d^3p_2}{(2\pi)^3}\left(2p_2^i+p^i\right)\,.
\ee
Note that we have disregarded the exponent $\exp({-iE(\vec p,\vec p_2) t})$ in the above expression. The reason for this is that as we shall see below the integral above is already first order in $\vec p$. Therefore, if we were to take the derivative with respect to $p^i$ of the energy in the exponent, it would come multiplied by a quantity linear in $\vec p$ and would vanish in the $\vec p \to 0$ limit. 

We evaluate the integral above in  spherical coordinates, aligning the $z$-axis with $\vec p$. Since the integral involves various approximations and gives a non-trivial result with important implications we present its full evaluation in App.~\ref{app:integral}. The end result for small $\vec p$ is
\be\label{integral_main}
\begin{split}
\int\frac{d^3p_2}{(2\pi)^3}\left(2\vec p_2+\vec p \, \right)=\frac{1}{6\pi^2}p_{\rm F}^3\vec p+\mathcal O(p^2)\,.
\end{split}
\ee
As for the right hand side of the theorem \eqref{theorem Tmn FL}, we use the fact that in the non-relativistic limit $\rho_{\text{FL}}\gg P_{\text{FL}}$ and evaluate 
\be\label{rhoFL}
\begin{split}
\rho_{\text{FL}}&=\langle\text{FL}|:T^{00}_{\text{N.R.}}(0):|\text{FL}\rangle=m\int\frac{d^3k\,d^3q}{(2\pi)^6}\langle \text{FL}|a^\dagger_{\vec k}a_{\vec q}|\text{FL}\rangle=m\int_{|\vec k|\leq p_{\rm F}}\frac{d^3k}{(2\pi)^3}\\
&=\frac{1}{6\pi^2}mp_{\rm F}^3\,.
\end{split}
\ee
By definition $\rho_{\rm FL}\equiv m\cdot \frac{N}{V}$, where $N$ is the total number of particles inside the Fermi surface. The above equation in turn implies $\int_{|\vec k|\leq p_{\rm F}}\frac{d^3k}{(2\pi)^3}=\frac{\rho_{\text{FL}}}{m} = \frac{N}{V}$, coinciding with the number density $n=N/V$. 
We further note that, due to the bosonic nature of the operators, the number density for our system of free bosons is half of that of a free Fermi gas, where an additional factor of $2$ arises due to the sum over spins \cite{Landau:1980mil}.
Inserting \eqref{integral_main} and \eqref{rhoFL} in the Goldstone theorem \eqref{theorem Tmn FL}, we see that both sides are equal and thus the theorem is satisfied for particle--hole excitations around the Fermi liquid ground state. 

\subsubsection*{Fermions}
The stress tensor for fermions can be derived from the action of free relativistic fermions 
\be
S=\int d^4 x\,\bar \psi(i\slashed\partial-m)\psi\,,
\ee
where $\bar \psi\equiv\psi^\dagger\gamma^0$, $\slashed\partial\equiv\gamma^\mu\partial_\mu$, and $\gamma^\mu$ are the standard gamma matrices. For the properly symmetrized stress tensor we obtain:
\be
T^{\mu\nu}(x)=\frac{i}{4}\left[\bar\psi\gamma^\mu\partial^\nu\psi-\partial^\nu\bar\psi\gamma^\mu\psi +(\mu\leftrightarrow\nu)\right]-\eta^{\mu\nu}\mathcal L\,,
\ee
where  $\mathcal L$ is the Lagrangian density. We then use the standard mode expansion for fermions:
\be\label{modes_ferm}
\psi=\int\frac{d^3p}{(2\pi)^3}\frac{1}{\sqrt{2\omega_p}}\sum_s\left(c^s_{\vec p} \, u^s_{\vec p} \, e^{-ip\cdot x}+d^{s\dagger}_{\vec p}v^s_{\vec p} \, e^{ip \cdot x}\right)\,,
\ee
where $c^s_{\vec p},c^{s\dagger}_{\vec p}$ and $d^s_{\vec p},d^{s\dagger}_{\vec p}$ are the creation and annihilation operators for particles and antiparticles,  respectively. The spinors $u^s_{\vec p}$ and $v^s_{\vec p}$ are the suitably normalized solutions of the Dirac equations satisfying $(\slashed p-m)u^s_{\vec p}=0$ and $(\slashed p+m)v^s_{\vec p}=0$. We give the full form of the solutions as well as any other notations used here in Appendix~\ref{app:free_fermions}. In the non-relativistic limit, we find for the normally ordered stress-energy tensor:
\be\label{T00_ferm}
:T^{00}_{\rm N.R.}(0):=\int\frac{d^3k}{(2\pi)^3}\frac{d^3q}{(2\pi)^3}\sum_s m\left(c^{s\dagger}_{\vec k}\,c^s_{\vec q}+d^{s\dagger}_{\vec k}d^s_{\vec q}\right)
\ee
and
\be\label{T0i_ferm}
\begin{split}
:T^{0i}_{\rm N.R.}(0):=\frac{1}{2}\int\frac{d^3k}{(2\pi)^3}\frac{d^3q}{(2\pi)^3}&\left[\sum_s \left(c^{s\dagger}_{\vec k}\,c^s_{\vec q}+d^{s\dagger}_{\vec k}d^s_{\vec q}\right)(k^i+q^i)\right]\\
+&\left.\frac{1}{2}\sum_{s,s'}\left(c^{s\dagger}_{\vec k}\,c^{s'}_{\vec q}+d^{s'\dagger}_{\vec k}d^{s}_{\vec q}\right)\xi^{s\dagger}\sigma^i\sigma^j(q^j-k^j)\xi^{s'}\right]\,,
\end{split}
\ee
where $\xi^1=\begin{pmatrix} 1\\0\end{pmatrix}$ and $\xi^2=\begin{pmatrix} 0\\1\end{pmatrix}$. The term on the second line of the above expression, when evaluated inside $\left\langle\text{FL}| \dots |\psi\right\rangle$, comes out proportional to $\vec p=\vec p_1-\vec p_2$ and thus is of higher order in $p$ and can be neglected. Thus, the relevant matrix elements $\mathcal T^{00}(\vec p,\vec p_2)$ and $\mathcal T^{0i}(\vec p,\vec p_2)$ coincide with \eqref{elements} of the bosonic case (when the contribution from antiparticles is disregarded). This similarly as before leads to 
\be
\lim_{\vec p\to 0} \sum_s\int \frac{d^3p_2}{(2\pi)^3}\, {\cal T}^{00}(\vec p \,,\vec p_2 ) \, {\cal T}^{0j} {}^*(\vec p \,,\vec p_2 ) + {\rm c.c.} =\frac{1}{3\pi^2}mp_{\rm F}^3p^i\,.
\ee
For the particle energy density we find similarly as in \eqref{rhoFL}:
\be
\begin{split}
\rho_{\text{FL}}=m\sum_s\int\frac{d^3k\,d^3q}{(2\pi)^6}\langle \text{FL}|c^{s\dagger}_{\vec k}c^s_{\vec q}|\text{FL}\rangle=2m\int_{|\vec k|\leq p_{\rm F}}\frac{d^3k}{(2\pi)^3}=\frac{1}{3\pi^2}mp_{\rm F}^3\,,
\end{split}
\ee
consistent with the energy density of the free Fermi gas \cite{Landau:1980mil}. Inserting the above results in the Goldstone theorem \eqref{theorem Tmn FL} we see that it is satisfied for the case of free fermions.

\subsubsection{Symmetry arguments}
There is however a simpler way of deriving the matrix elements of $T^{\mu\nu}_{\rm free}$ at position $x=0$ needed for our Goldstone theorem for the free theory. In particular, the stress-tensor of a free theory can be written in terms quadratic in creation and annihilation operators as
\be
\begin{split}\label{ansatz}
&T^{\mu\nu}_{\rm free}(0)=\int \frac{d^3k}{(2\pi)^3}\frac{d^3q}{(2\pi)^3}b^{\dagger}_{\vec k}b_{\vec q}\,F^{\mu\nu}(\vec k,\vec q\,)\,,
\end{split}
\ee
where $(b^\dagger_{\vec k},b_{\vec k})$ can be either bosonic creation and annihilation operators, \emph{i.e.}~$a^\dagger_{\vec k},a_{\vec k}$ with $[a_{\vec k},a^\dagger_{\vec q}]=(2\pi)^3\delta^{(3)}(\vec k-\vec q \, )$, or fermionic ones, \emph{i.e.}~$c^{s \dagger}_{\vec k},c^s_{\vec k}$ with $\{c^s_{\vec k},c^{s' \dagger}_{\vec q}\}=(2\pi)^3\delta^{ss'}\delta^{(3)}(\vec k-\vec q \,)$, and $F^{\mu\nu}(\vec k,\vec q\,)$ are some functions to be characterized later. Henceforth we shall drop the vector signs from the arguments of $F^{\mu\nu}$ in order not to clutter the notations any further. Note that on the Fermi Liquid ground state we obtain
\be
 \langle \text{FL}|T^{\mu\nu}_{\rm free}(0) |\text{FL}\rangle=\int _{|\vec k|\leq p_{\rm F}}\frac{d^3k}{(2\pi)^3}F^{\mu\nu}(k,k)\,,
\ee
where we have used \eqref{elements}. In fact, it is straightforward to check that \eqref{elements} holds also for fermions. We thus use for the matrix elements $\langle\text{FL}|b_{\vec p}^{\dagger} \, b_{\vec k}  |\text{FL}\rangle=\left.(2\pi)^3\delta^{(3)}(\vec p-\vec k)\right|_{|\vec p \, |,|\vec k|\leq p_{\rm F}}$. Let us remark that, since we are considering the case of free particles, it is well justified to extend the integral above to the inside of the Fermi sphere. When considering interactions we shall keep in mind that the theory is weakly coupled only in the momentum band close to the Fermi surface, as shown in Fig.~\ref{Fig:fermi} [left].

On can then show that the matrix elements \eqref{matrix_elem} needed for the theorem can be expressed as
\be\label{state_relation_0}
\begin{split}
\langle \text{FL}|T^{\mu\nu}_{\rm free}(0)|\psi\rangle
&=\langle \mathbb 1_{\vec p_2}|T^{\mu\nu}_{\rm free}(0)|\mathbb 1_{\vec p_1}\rangle\pm\langle\text{FL}|\{[T^{\mu\nu}_{\rm free}(0),b_{\vec p_2}]\}b^\dagger_{\vec p_1}|\text{FL}\rangle\\
& \quad\mp(2\pi)^3\delta^{(3)}(\vec p_1-\vec p_2)\langle \text{FL}|T^{\mu\nu}_{\rm free}(0) |\text{FL}\rangle\,,
\end{split}
\ee
where, as before, the particle--hole state is $|\psi\,\rangle\equiv {b^{\dagger}_{\vec p_1}}\,b_{\vec p_2}|\text{FL}\rangle$ and we have defined the single particle states as $\left|\mathbb 1_{\vec p_1}\right\rangle\equiv b^{\dagger}_{\vec p_1}\left|\text{FL}\right\rangle$. The upper sign and the commutator are for the bosonic case, while the lower sign and the anti-commutator are for the fermionic case. To derive the above equation, we have used that for any operator $\mathcal O$ the following relationship holds:
\be
\mathcal O b^\dagger_{\vec p_1}b_{\vec p_2}=\mathcal O\{[b^\dagger_{\vec p_1},b_{\vec p_2}]\}\pm\{\left[\mathcal O,b_{\vec p_2}\right]\}b^\dagger_{\vec p_1}+b_{p_2}\mathcal O b^\dagger_{\vec p_1}\,.
\ee
where the $\pm$ and $\{[\dots]\}$ notation is the same as above. 

In order to deal with the second term in \eqref{state_relation_0}, we further use that for fermions
\be\label{commutator_p2}
\{c^\dagger_{\vec k}c_{\vec q},c_{\vec p_2}\}c^\dagger_{\vec p_1}=2c_{\vec p_2}(c^\dagger_{\vec k}c_{\vec q})c^\dagger_{\vec p_1}-\{c^\dagger_{\vec k},c_{\vec p_2}\}c_{\vec q}c^\dagger_{\vec p_1}\,,
\ee
while for bosons
\be\label{bosons_easy}
[a^\dagger_{\vec k}a_{\vec q},a_{\vec p_2}]a^\dagger_{\vec p_1}=[a^\dagger_{\vec k},a_{\vec p_2}]a_{\vec q}a^\dagger_{\vec p_1}\,.
\ee
Inserting this into \eqref{state_relation_0}  leads to 
\be\label{state_relation}
\begin{split}
{\cal T}^{\mu\nu}(\vec p,\vec p_2)\equiv\langle \text{FL}|T^{\mu\nu}_{\rm free}(0)|\psi\rangle
&=\pm \langle \mathbb 1_{\vec p_2}|T^{\mu\nu}_{\rm free}(0)|\mathbb 1_{\vec p_1}\rangle\mp F^{\mu\nu}(p_2,p_1)\\
& \quad\mp(2\pi)^3\delta^{(3)}(\vec p_1-\vec p_2)\langle \text{FL}|T^{\mu\nu}_{\rm free}(0) |\text{FL}\rangle\,,
\end{split}
\ee
where again the upper sign is for bosons and the lower sign is for fermions. We shall generalize these relationships to arbitrary operators when discussing the interacting theory in Section~\ref{sec:interactions}. Let us emphasize that in the above equation we are still only interested in the situation when $|\vec p_1|>p_F$ and $|\vec p_2|\leq p_F$, and in particular $\vec p_1\neq \vec p_2$. In this case the last term in \eqref{state_relation} is absent while the matrix element $ \langle \mathbb 1_{\vec p_2}|T^{\mu\nu}_{\rm free}(0)|\mathbb 1_{\vec p_1}\rangle$ vanishes for fermions (because $\langle\text{FL}| c^s_{\vec p_2}=0$) and is non-zero for bosons. We shall come back to this in a moment. 

The benefit of equation \eqref{state_relation} is that it relates the FL--(particle-hole) matrix elements to particle--particle matrix elements. This turns out to be particularly handy in the situation when $\vec p_1\to \vec p_2$. In that case one can use the fact that for any single particle state with mass $m$ and momentum $\vec k$ above the Fermi surface, \emph{i.e.}~with $|\vec k \, |>p_F$, the stress-energy tensor components in the non-relativistic limit have to satisfy\footnote{Had we defined the single-particle states on a Lorentz invariant ground state, \emph{i.e.}~$\left|\vec p\,\right\rangle\equiv b^{\dagger}_{\vec p}\left|0\right\rangle$, we could use the discussion of subsection~\ref{sec:massive_part} to justify this. In particular, from \eqref{Tmn_single_part} it follows that when evaluated on a single-particle state the stress energy tensor has to obey $\left\langle \vec p\,\right|T^{00}_{\rm free}(0)\left|\vec p\,\right\rangle=p^0\approx m$ and $\left\langle \vec p\,\right|T^{0i}_{\rm free}(0)\left|\vec p\,\right\rangle=p^i$.} 
\begin{align}\label{match_00}
&\left\langle \mathbb 1_{\vec k} \right|T^{00}_{\rm free}(0)\left|\mathbb 1_{\vec k} \right\rangle=\rho_{\rm FL}V+m\,,\\\label{match_0i}
&\left\langle \mathbb 1_{\vec k} \right|T^{0i}_{\rm free}(0)\left|\mathbb 1_{\vec k} \right\rangle=k^i\,,
\end{align}
where $\rho_{\rm FL}$ is the energy density of the Fermi Liquid in its ground state as defined in \eqref{fermi_gs}. Similar relations also hold in the case of single \emph{hole} states defined as $|\bar{\mathbb 1}_{\vec k}\rangle\equiv b_{\vec k}|\text{FL}\rangle$ and 
 $\langle\bar{\mathbb 1}_{\vec k}|\equiv \langle\text{FL}|b^{\dagger}_{\vec k}$ with $|\vec k|\leq p_{\rm F}$ when in the non-relativistic limit we require
\begin{align}\label{match00h}
&\left\langle \bar{\mathbb 1}_{\vec k}\right|T^{00}_{\rm free}(0)\left|\bar{\mathbb 1}_{\vec k}\right\rangle=\rho_{\rm FL}V-m\,,\\\label{match0ih}
&\left\langle \bar{\mathbb 1}_{\vec k}\right|T^{0i}_{\rm free}(0)\left|\bar{\mathbb 1}_{\vec k}\right\rangle=-k^i\,.
\end{align}
This then completely fixes the equal-momentum components of the function $F^{\mu\nu}$:
\be\label{Fs3}
F^{00}(k,k)=m\,,\qquad F^{0i}(k,k)=k^i\,,
\ee
and, importantly, the functions $F^{\mu\nu}(k,k)$ are continuous at the Fermi surface. 
 To show how this works out in full generality we reintroduce the spin index for the operators $b^s, b^{s \dagger}$ and functions $F^{\mu\nu}_{ss'}(k,q)$ and evaluate the matrix elements $\langle \mathbb 1_{\vec p_2}^{s_2}|T^{\mu\nu}_{\rm free}(0)|\mathbb 1_{\vec p_1}^{s_1}\rangle$ in Eq.~\eqref{state_relation} for  various configurations of momenta $\vec p_1$ and $\vec p_2$ in Appendix~\ref{sec:elements}.  
 
 With all the above relationships at hand we have now all the necessary ingredients to evaluate the matrix elements ${\cal T}^{00}(\vec p \,,\vec p_2)$ and ${\cal T}^{0i}(\vec p \,,\vec p_2)$ needed for the Goldstone theorem \eqref{theorem Tmn FL}. Starting from the relation \eqref{state_relation} between the matrix elements evaluated on a particle--hole state and on the single-particle states, the next steps differ slightly depending on whether we are considering a bosonic or fermionic Fermi liquid state.

\subsubsection*{Bosons}\label{symmetry free bosons}

Using the result \eqref{part_part} for the single particle matrix elements for $|\vec p_1|>p_{\rm F}$, $|\vec p_2|\leq p_{\rm F}$ and $\vec p_1\neq\vec p_2$, the relation \eqref{state_relation} for bosons becomes 
\be\label{calTmunu}
\begin{split}
{\cal T}^{\mu\nu}(\vec p,\vec p_2)&= \langle \mathbb 1_{\vec p_2}|T^{\mu\nu}_{\rm free}(0)|\mathbb 1_{\vec p_1}\rangle-F^{\mu\nu}(p_2,p_1)\,\\
&=2F^{\mu\nu}(p_2,p_1)-F^{\mu\nu}(p_2,p_1)\\
&=F^{\mu\nu}(p_2,p_1)\,.
\end{split}
\ee
Let us now evaluate the product ${\cal T}^{00}(\vec p \,,\vec p_2 ) \, {\cal T}^{0j} {}^*(\vec p \,,\vec p_2 ) $ in the limit $\vec p\to 0$ (or equivalently $\vec p_1\to \vec p_2$). We first note that we have established in Appendix~\ref{app:integral} that the integration measure in the Goldstone theorem \eqref{theorem Tmn FL} is itself already linear in $p^i$. Indeed, we have found
\be
\lim_{\vec p\to 0} \int\frac{d^3p_2}{(2\pi)^3}=\frac{p_{\rm F}^2}{(2\pi)^3}\int_0^{2\pi} d\varphi \,\int_{0}^1d\cos \theta' \int _{0}^{-p\cos\theta'}d\delta p_2+\mathcal O(p^2)=-\frac{p_{\rm F}^2p}{4\pi^2}\int_0^1d\cos\theta'\,\cos\theta'\,.
\ee
Hence, for our Goldstone theorem we are only interested in the matrix elements ${\cal T}^{\mu\nu}(\vec p,\vec p_2)$ in \eqref{calTmunu} up to the zeroth order in $p$. 
In particular, expanding in the limit $\vec p_1\to \vec p_2$ we get\footnote{Note that we could have equivalently chosen to take the limit $\vec p\to 0$ by sending $\vec p_2\to \vec p_1$. In that case we would need to evaluate $F^{\mu\nu}(p_1,p_1)$ with $|\vec p_1|>p_{\rm F}$. As we have shown in Appendix~\ref{sec:elements} the functions $F^{\mu\nu}(k,k)$ are continuous at the Fermi surface and thus the result would remain unchanged.}
\be\label{elem_limit}
\lim_{\substack{\vec p_1\to \vec p_2\\ \vec p_1\neq\vec p_2}}F^{\mu\nu}(p_2,p_1)=F^{\mu\nu}(p_2,p_2)+\left.\partial_{p_1^i}F^{\mu\nu}(p_2,p_1)\right|_{\vec p_1=\vec p_2}\cdot p^i+\mathcal O(p^2)\,
\ee
and only need to retain the leading term for which we have found earlier that $F^{00}(p_2,p_2)$, $F^{0i}(p_2,p_2)$ obey \eqref{Fs3}. Hence, we do not need any additional information to find
\be
\lim_{\vec p\to 0}{\cal T}^{00}(\vec p \,,\vec p_2 ) \, {\cal T}^{0j} {}^*(\vec p \,,\vec p_2 ) =\lim_{\vec p_1\to\vec p_2}F^{00}(p_2,p_1)F^{0j}{}^*(p_2,p_1)=mp_2^j=mp_{\rm F}\hat p_2^j+\mathcal O(p)\,.
\ee
Substituting this in \eqref{theorem Tmn FL} we get for the Goldstone theorem
\be
\frac{mp_{\rm F}^3}{6\pi^2}\delta^{ij}=\rho_{\rm FL}\delta^{ij}\,.
\ee
Using the expression for the energy density in \eqref{rhoFL} we see again that the theorem is satisfied.

\subsubsection*{Fermions}
The analysis is different for fermionic Fermi liquids. The equation \eqref{state_relation} becomes:
\be\label{argument_1}
\underbrace{\langle \text{FL}|T^{\mu\nu}_{\rm free}(0)c^\dagger_{\vec p_1}c_{\vec p_2}|\text{FL}\rangle}_{=0,\,|\vec p_2|>p_{\rm F}}+ \underbrace{\langle \mathbb 1_{\vec p_2}|T^{\mu\nu}_{\rm free}(0)|\mathbb 1_{\vec p_1}\rangle}_{=0,\,|\vec p_2|\leq p_{\rm F}}=F^{\mu\nu}(p_2,p_1)\,,
\ee
where we have used that $|\vec p_1|>p_{\rm F}$ while keeping arbitrary values of $\vec p_2$. This implies, that
\be
\begin{cases}
&\langle \text{FL}|T^{\mu\nu}_{\rm free}(0)c^\dagger_{\vec p_1}c_{\vec p_2}|\text{FL}\rangle=F^{\mu\nu}(p_2,p_1)\,,\qquad |\vec p_2|\leq p_{\rm F}\,,\\
    &\langle \mathbb 1_{\vec p_2}|T^{\mu\nu}_{\rm free}(0)|\mathbb 1_{\vec p_1}\rangle=F^{\mu\nu}(p_2,p_1)\,,\qquad|\vec p_2|>p_{\rm F}\,.
\end{cases}
\ee
Since the function $F^{\mu\nu}(p_2,p_1)$ is continuous in $p_2\approx p_{\rm F}$ (\emph{i.e.} we have not in any way incorporated our knowledge of the existence of the Fermi surface in the parametrization \eqref{ansatz}), this means
\be
\lim_{\vec p_2\to \vec p_{{\rm F}-}} \langle \text{FL}|T^{\mu\nu}_{\rm free}(0)c^\dagger_{\vec p_1}c_{\vec p_2}|\text{FL}\rangle =\lim_{\vec p_2\to \vec p_{{\rm F}+}}\langle \mathbb 1_{\vec p_2}|T^{\mu\nu}_{\rm free}(0)|\mathbb 1_{\vec p_1}\rangle\,,
\ee
where we use the standard notations $\vec p_2\to \vec p_{{\rm F}-}$ meaning that $\vec p_2$ approaches $\vec p_{\rm F}$ from below (and similarly for $\vec p_2\to \vec p_{{\rm F}+}$). For the matrix elements needed for the Goldstone theorem we thus obtain for $|\vec p_1|>p_{\rm F}$, $|\vec p_2|\leq p_{\rm F}$ and $\vec p\equiv \vec p_1-\vec p_2$
\be
\begin{split}
\lim_{\substack{\vec p\to 0}}{\cal T}^{00}(\vec p \,,\vec p_2 ) =\lim_{\substack{\vec p_1\to \vec p_2,\\\vec p_2\to \vec p_{\rm F+}}}\left\langle \mathbb 1_{\vec p_2} ^{s_2}\right|T^{00}_{\rm free}(0)\left|\mathbb 1_{\vec p_1} ^{s_1}\right\rangle=\left.\left\langle \mathbb 1_{\vec p} ^s\right|T^{00}_{\rm free}(0)\left|\mathbb 1_{\vec p} ^s\right\rangle\right|_{|\vec p|>p_{\rm F}}-\rho_{\rm FL}V=m\,,
\end{split}
\ee
where in the last equality we have used the requirements \eqref{match_00} and \eqref{match_0i} in the non-relativistic limit. In the above expression we have also subtracted the contribution $\rho_{\rm FL} V$ from the equal-momenta single particle matrix elements. The reason for that is that we are considering the limit when $\vec p_1\to \vec p_2$, however, keeping in mind that $\vec p_1\neq \vec p_2$.  Similarly we obtain $\lim_{\substack{\vec p\to 0}}{\cal T}^{0j}(\vec p \,,\vec p_2 )= p^j_2$. 

For the right hand side of the theorem we need to evaluate $\rho_{\rm FL}$. Similarly as for the bosons in \eqref{rhoFL} we find
\be\label{rhoFL_free_ferm}
\begin{split}
\rho_{\text{FL}}&=\langle\text{FL}|T^{00}_{\text{free}}(0)|\text{FL}\rangle=\sum_s m\int_{|\vec k|\leq p_{\rm F}}\frac{d^3k}{(2\pi)^3}=\frac{1}{3\pi^2}mp_{\rm F}^3\,.
\end{split}
\ee
Combining this with the findings above we see that the Goldstone theorem \eqref{theorem Tmn FL} is satisfied in the case of free fermions. 

\subsection{Interactions}\label{sec:interactions}
In this subsection we consider the stress-energy tensor of a fully interacting theory. This means that instead of the free stress-energy tensor given in \eqref{ansatz} we shall now parametrize it as \be\label{ansatzTmn_int}
T^{\mu\nu}_{\rm interacting}(0)=\sum^\infty_{n=1} \int \prod_{i,j=1}^n\frac{d^3k_j}{(2\pi)^3}\frac{d^3q_i}{(2\pi)^3}\mathcal O^+_n\mathcal O^-_n\,F_n^{\mu\nu}(\vec k_1,\dots ,\vec k_n;,\vec q_1,\dots,\vec q_n)\,,
\ee
where we have extended the function $F_n^{\mu\nu}$ of \eqref{ansatz} to multiple arguments and have introduced the shorthand notation
\be
\mathcal O^+_n=\prod_{j=1}^nb^\dagger_{\vec k_j}\,,\qquad \mathcal O^-_n=\prod_{i=1}^nb_{\vec q_i}\,.
\ee
The $n=1$ term in the sum corresponds to the free case discussed above. The $n>1$ terms correspond to interactions. In fact, in Fermi liquid theory only certain $n=2$ terms survive when the quasi-particles' momenta approach the Fermi surface. Specifically, for a 2-2 scattering process the only kinematically allowed configurations are the ones with back-to-back incoming momenta and  forward scattering. All higher order terms are, in the renormalization group sense, irrelevant \cite{Polchinski:1992ed}.

\subsubsection{Fermions}
In the case of fermions, by using the identities
\be
\begin{split}
&c_{\vec p_2}\,\mathcal O^+_n=\sum_{j=1}^n(-1)^{j-1}(2\pi)^3\delta^{(3)}(\vec k_j-\vec p_2)c^\dagger_{\vec k_1}\dots c^\dagger_{\vec k_{j-1}}c^\dagger_{\vec k_{j+1}}\dots c^\dagger_{\vec k_n}+(-1)^n\mathcal O^+_n\,c_{\vec p_2}\,,\\
&\mathcal O^-_n\,c^\dagger_{\vec p_1}=\sum_{i=1}^n(-1)^{n-i}(2\pi)^3\delta^{(3)}(\vec q_j-\vec p_1)c_{\vec q_1}\dots c_{\vec q_{i-1}}c_{\vec q_{i+1}}\dots c_{\vec q_n}+(-1)^nc^\dagger_{\vec p_1}\,\mathcal O^-_n
\end{split}
\ee
one can show
\be\label{state_relation_2}
\begin{split}
 \langle \text{FL}|T^{\mu\nu}_{\rm interacting}(0)c^\dagger_{\vec p_1}c_{\vec p_2}|\text{FL}\rangle&= \langle \text{FL}|c_{\vec p_2}T^{\mu\nu}_{\rm int}(0)c^\dagger_{\vec p_1}|\text{FL}\rangle-\langle\text{FL}|\{T^{\mu\nu}_{\text{int}}(0),c_{\vec p_2}\}c^\dagger_{\vec p_1}|\text{FL}\rangle\\
 &+(2\pi)^3\delta^{(3)}(\vec p_1-\vec p_2)\left\langle\text{FL}|T^{\mu\nu}_{\text{int}}(0)|\text{FL}\right\rangle \,.
\end{split}
\ee
This is the generalization of \eqref{state_relation_0} for fermions. We can also generalize the relationship \eqref{commutator_p2} as
\be
\{\mathcal O^+_n\mathcal O^-_n,c_{\vec p_2}\}c^\dagger_{\vec p_1}=2c_{\vec p_2}\mathcal O^+_n\mathcal O^-_nc^\dagger_{\vec p_1}-\sum_{j=1}^n(-1)^{j-1}\{c_{\vec p_2},c^\dagger_{\vec k_j}\}c^\dagger_{\vec k_1}\dots c^\dagger_{\vec k_{j-1}}c^\dagger_{\vec k_{j+1}}\dots c^\dagger_{\vec k_n}\mathcal O^-_nc^\dagger_{\vec p_1}\,.
\ee
Indeed, we recover \eqref{commutator_p2} for $n=1$. 
This leads to 
\be\label{state_relation_3}
\begin{split}
\langle \text{FL}|&T^{\mu\nu}_{\rm int}(0)|\psi\rangle+ \langle \mathbb 1_{\vec p_2}|T^{\mu\nu}_{\rm int}(0)|\mathbb 1_{\vec p_1}\rangle=(2\pi)^3\delta^{(3)}(\vec p_1-\vec p_2)\langle \text{FL}|T^{\mu\nu}_{\rm int}(0) |\text{FL}\rangle\\
&+\sum_{n=1}^\infty\sum_{j=1}^n(-1)^{j-1}(2\pi)^3\int \prod_{i,l=1}^n\frac{d^3k_l}{(2\pi)^3}\frac{d^3q_i}{(2\pi)^3}\delta^{(3)}(\vec p_2-\vec k_j)\langle\text{FL}|c^\dagger_{\vec k_1}\dots c^\dagger_{\vec k_{j-1}}c^\dagger_{\vec k_{j+1}}\dots c^\dagger_{\vec k_n}\mathcal O^-_nc^\dagger_{\vec p_1}|\text{FL}\rangle\\
&\qquad\times F_n^{\mu\nu}(k_1,\dots ,k_n;q_1,\dots,q_n)\,,
\end{split}
\ee
where we have put the relation \eqref{state_relation_2} in a form such that the operator $c_{\vec p_2}$ does not appear explicitly in the last term anymore. As before, we have also dropped the vector signs on the arguments of $F^{\mu\nu}_n$. 
The last term in the above expression can be simplified further by using the fact that the function $F_n^{\mu\nu}(k_1,\dots,k_n; q_1,\dots,q_n)$ is antisymmetric in $k$'s and $q$'s, so that
\be
F_n^{\mu\nu}(k_1,\dots,k_n; q_1,\dots,q_n)=\frac{1}{n!}\varepsilon_{i_1\dots i_n}F_n^{\mu\nu}(k_{i_1},\dots,k_{i_n};q_1,\dots,q_n)\,,
\ee
and similar for $q$'s. Another useful identity is 
\be
\langle\text{FL}|c^\dagger_{\vec k_n}\dots c^\dagger_{\vec k_1}\,c_{\vec q_1}\dots c_{\vec q_n}|\text{FL}\rangle=\varepsilon_{i_1\dots i_n}\delta_{k_{i_1}q_1}\delta_{k_{i_2}q_2}\dots\delta_{k_{i_n}q_n}\,.
\ee
We then find for the last term in \eqref{state_relation_3}
\be\label{gen_fermions}
\begin{split}
\sum_{j=1}^n(-1)^{j-1}\int &\prod_{i,l=1}^n\frac{d^3k_l}{(2\pi)^3}\frac{d^3q_i}{(2\pi)^3}\delta_{p_2k_j}\langle\text{FL}|c^\dagger_{\vec k_1}\dots c^\dagger_{\vec k_{j-1}}c^\dagger_{\vec k_{j+1}}\dots c^\dagger_{\vec k_n}\mathcal O^-_nc^\dagger_{\vec p_1} |\text{FL}\rangle F_n^{\mu\nu}(k_1,\dots ,k_n;q_1,\dots,q_n)\\
&=(-1)^{n-1}nn!\int \frac{d^3l_1}{(2\pi)^3}\dots\frac{d^3l_{n-1}}{(2\pi)^3}F_n^{\mu\nu}(p_2,l_1,\dots l_{n-1};p_1,l_1,\dots,l_{n-1})\\
&\equiv (-1)^{n-1}nn!\tilde F_n^{\mu\nu}(p_2,p_1)\,,
\end{split}
\ee
where in the last step we have defined a function $\tilde F_n^{\mu\nu}(p_2,p_1)$, analogous to the $F^{\mu\nu}(p_2,p_1)$ in the case of free fermions. Indeed, for $n=1$ we recover from the definition above that $\tilde F_1^{\mu\nu}(p_2,p_1)=F^{\mu\nu}(p_2,p_1)$. The equation \eqref{state_relation_3} now takes the form
\be\label{state_relation_4}
\begin{split}
\langle \text{FL}|T^{\mu\nu}_{\rm int}(0)|\psi\rangle+ &\langle \mathbb 1_{\vec p_2}|T^{\mu\nu}_{\rm int}(0)|\mathbb 1_{\vec p_1}\rangle=\delta_{p_1p_2}\langle \text{FL}|T^{\mu\nu}_{\rm int}(0) |\text{FL}\rangle+\sum_{n=1}^\infty (-1)^{n-1}nn!\tilde F_n^{\mu\nu}(p_2,p_1)\,,
\end{split}
\ee
matching the equation \eqref{state_relation} for fermions. Here we only care that this last term, $\tilde F_n^{\mu\nu}(p_2,p_1)$ is a function, continuous in $\vec p_2$. We have shown in the previous subsection that this is true for free fermions and we have no reason to doubt that it also holds for interacting fermions---the stress--energy tensor in \eqref{ansatzTmn_int} should not know of the existence of the Fermi surface. Following the same argument as in the case of free fermions (\emph{i.e.}~a stress--energy tensor quadratic in creation and annihilation operators) we conclude that also here 
\be
\lim_{\vec p_2\to \vec p_{F-}} \langle \text{FL}|T^{\mu\nu}_{\rm int}(0)c^\dagger_{\vec p_1}c_{\vec p_2}|\text{FL}\rangle =\lim_{\vec p_2\to \vec p_{F+}}\langle \mathbb 1_{\vec p_2}|T^{\mu\nu}_{\rm int}(0)|\mathbb 1_{\vec p_1}\rangle\,.
\ee
The single-particle elements on the r.h.s. can be evaluated in the non-relativistic limit giving:
\be
\lim_{\substack{\vec p\to 0}}{\cal T}^{00}(\vec p \,,\vec p_2 ) =m\,,\qquad\lim_{\substack{\vec p\to 0}}{\cal T}^{0i}(\vec p \,,\vec p_2 )=p^i_2\,.
\ee
To prove the theorem one can then proceed similarly as in the case of free fermions by expressing the mass $m$ and the energy density $\rho_{\rm FL}$ in terms of the functions $\tilde F^{00}_n$, similarly as was done in \eqref{Fs3} and \eqref{rhoFL_free_ferm}. In particular, we find
\be\label{exp_value_FL}
\langle \text{FL}|T^{\mu\nu}_{\rm int}(0) |\text{FL}\rangle=\sum_s\int\frac{d^3k}{(2\pi)^3}\sum_{n=1}^\infty n!\tilde F_{n,ss}^{\mu\nu}(k,k)\,,
\ee
where we have reintroduced the spin indices in order to get the correct factors of $2$ arising from summing over the spin states in a Fermi gas. For the single \emph{hole} matrix elements we find in turn
\be\label{single_hole_int}
\left\langle \bar{\mathbb 1}_{\vec p_2}^{s_2}\right|T^{\mu\nu}_{\rm int}(0)\left|\bar{\mathbb 1}_{\vec p_1}^{s_1}\right\rangle=-\sum_{n=1}^\infty nn!\tilde F^{\mu\nu}_{n,s_1s_2}(p_1,p_2)+\delta_{p_1p_2}\langle \text{FL}|T^{\mu\nu}_{\rm int}(0) |\text{FL}\rangle\,.
\ee
This is very similar to the result obtained in the free theory --- it reduces to \eqref{single_hole} for $n=1$. We would now like to map this to the non-relativistic relationships \eqref{match00h} and \eqref{match0ih} in the case when both momenta are equal. We obtain
\begin{align}\label{mass_int}
V\rho_{\rm FL}-m&=V\langle \text{FL}|T^{00}_{\rm int}(0) |\text{FL}\rangle-\sum_{n=1}^\infty nn!\tilde F^{00}_{n,ss}(p,p)\,,\\
-p^i&=-\sum_{n=1}^\infty nn!\tilde F^{0i}_{n,ss}(p,p)\,.
\end{align}
The first equality above naturally identifies $\rho_{\rm FL}=\langle\text{FL}|T^{00}_{\rm int}(0)|\text{FL}\rangle$. Recall that in the free theory we were further able to express the energy density as $\rho_{\rm FL}=m\cdot \frac{N}{V}=m\sum_s\int_{|\vec k|\leq p_{\rm F}}\frac{d^3k}{(2\pi)^3}$. This seems to be problematic here if we were to identify the entire second term on the r.h.s. of  \eqref{mass_int} with the mass. Indeed, note the additional factor of $n$ in comparison to the Fermi liquid state expectation value \eqref{exp_value_FL}. In order to satisfy the equations above we thus have to require that in the non-relativistic limit:
\begin{align}\label{conditions}
\tilde F^{00}_{1,ss}(p,p)=m\,,\qquad \tilde F^{0i}_{1,ss}(p,p)=p^i\,,\qquad \tilde F^{0\mu}_{n\neq 1,ss}(p,p)=0\,.
\end{align}
In other words, we require that both the mass and momentum of these quasi-particles are set by the non-relativistic limit of the free stress-energy tensor and do not get renormalized by interactions. We can then further identify
\be\label{rhoFL_int}
\rho_{\rm FL}=\langle \text{FL}|T^{00}_{\rm int}(0) |\text{FL}\rangle=\sum_{n=1}^\infty n!\sum_s\int\frac{d^3k}{(2\pi)^3}\tilde F_{n,ss}^{00}(k,k)=m\sum_s\int_{|\vec k|\leq p_{F}}\frac{d^3k}{(2\pi)^3}\,.
\ee
With these identifications at hand we can now work out both sides of our Goldstone theorem \eqref{theorem Tmn FL}  and see that it is again satisfied. 

Finally, let us further remark that, as a result of our non-relativistic approximation, the expression \eqref{rhoFL_int}  for the mass density coincides with the mass density \eqref{rhoFL_free_ferm} of the free fermion gas. This approximation thus reduces to the original argument of Landau \cite{Lifshitz}, stating that an interacting Fermi liquid has the same relationship between number density and $p_F$ as a free Fermi gas. We find however that this can only be true if the functions $F^{\mu\nu}_n$ describing the fermion interactions satisfy the conditions \eqref{conditions}. In the most relevant case of $n=2$, these conditions read $\tilde F^{0\mu}_{2}(\vec p,\vec p\,)=\int \frac{d^3k}{(2\pi)^3}F^{0\mu}_2(\vec p,\vec k;\vec p,\vec k\,)=0$. The  structure of the momentum arguments  suggests that only the  2--2 scattering processes with equal ingoing and outgoing momenta (\emph{i.e.} $\vec p_{in,1}=\vec p_{out, 1}=\vec p\,$) contribute to  mass renormalization away from the non-relativistic limit. This corresponds to one of the special kinematic configurations of Fermi liquids---forward scattering.
\subsubsection{Bosons}\label{symmetry interacting bosons}
To work out the matrix elements needed for the Goldstone theorem in \eqref{theorem Tmn} for interacting bosons, we shall follow the same steps as for the fermions analyzed in the previous subsection. One can show that
\be\label{state_relation_5}
\begin{split}
 \langle \text{FL}|T^{\mu\nu}_{\rm interacting}(0)a^\dagger_{\vec p_1}a_{\vec p_2}|\text{FL}\rangle&= \langle \text{FL}|a_{\vec p_2}T^{\mu\nu}_{\rm int}(0)a^\dagger_{\vec p_1}|\text{FL}\rangle+\langle\text{FL}|[T^{\mu\nu}_{\text{int}}(0),a_{\vec p_2}]a^\dagger_{\vec p_1}|\text{FL}\rangle\\
 &-(2\pi)^3\delta^{(3)}(\vec p_1-\vec p_2)\left\langle\text{FL}|T^{\mu\nu}_{\text{int}}(0)|\text{FL}\right\rangle \,,
\end{split}
\ee
coinciding with the relation \eqref{state_relation_0} derived in the case of free bosons. As before, one needs to deal with the second term in the above expression. In quadratic case we used \eqref{bosons_easy} to obtain \eqref{state_relation}. And then as the second step we used the expression \eqref{part_part} to arrive at the final result in Eq.~\eqref{calTmunu}.  While it is still possible to generalize \eqref{bosons_easy} to  
\be
[\mathcal O^+_n\mathcal O^-_n,a_{\vec p_2}]a^\dagger_{\vec p_1}=-\sum_{j=1}^n[a_{\vec p_2},a^\dagger_{\vec k_j}]a^\dagger_{\vec k_1}\dots a^\dagger_{\vec k_{j-1}}a^\dagger_{\vec k_{j+1}}\dots a^\dagger_{\vec k_n}\mathcal O^-_na^\dagger_{\vec p_1}\,,
\ee
after inserting in \eqref{state_relation_5} one is left with a sum of integrals over internal momenta of matrix elements of the form
\be
\langle\text{FL}|a^\dagger_{\vec k_a}\dots a^\dagger_{\vec k_n}\,a_{\vec q_1}\dots a_{\vec q_n}|\text{FL}\rangle=\sum_{i_n\in \mathbb S_n}\delta_{k_{i_1}q_1}\delta_{k_{i_2}q_2}\dots\delta_{k_{i_n}q_n}-\frac{n!\delta_{n> 1}}{V^{n-1}}\delta_{k_1k_2}\delta_{k_2k_3}\dots\delta_{k_nq_1}\dots\delta_{q_{n-1}q_n}\,.
\ee
Here $\mathbb S_n$ denotes the group of all possible permutations of $n$ symbols and we use $\delta_{n>1}$ to emphasize that this last term is only added in the case when $n>1$. It is these last terms appearing for the cases of coincidental internal momenta that (perhaps not unexpectedly) are problematic. In particular, we have checked explicitly that a simplification (such as what is occurring for fermions) allowing to simplify \eqref{state_relation_5} to something similar to \eqref{state_relation_4} is not happening for bosons. Physically this was to be expected because naturally bosons are not obeying Pauli exclusion principle (nor Fermi statistics, as a consequence), needed to construct the Fermi liquid ground state.

\section{Concluding remarks and outlook}\label{sec:conclusions}
Goldstone theorems provide essential information about the low-energy excitations in systems with spontaneous symmetry breaking. In this work we have focused on systems with broken invariance under Lorentz boosts. These are known to exhibit peculiar properties --- in most cases the boost Goldstone bosons are non-dynamical and are absorbed in the Goldstones of some other broken spacetime symmetries. A particularly curious example is the case of framids --- systems in which there are no other broken symmetries and thus no other Goldstone bosons that could govern the low-energy dynamics. In such a situation we do expect to see the boost Goldstones, even if due to the peculiar properties of their stress-energy tensor (satisfying $\rho+p=0$) such systems do not appear to be realized in laboratory. On the other hand there is a prominent condensed matter example in which indeed no other symmetries apart from the Lorentz boosts are broken --- the Fermi liquids. These are strongly coupled phases of interacting fermions at temperatures so low that the dynamics are completely determined by quantum effects. As such, even though the material is in a liquid state its dynamics is not described in terms of the usual collective excitations of fluid dynamics, the phonons. 

With this in mind, we have derived from  first principles a boost Goldstone theorem in terms of a generic order parameter, which can be conveniently chosen in a way that is most suited for the medium under consideration.
We have shown how the theorem is satisfied by standard examples of phases of matter with broken boost invariance---framids and superfluids. Considering a  toy example of a reference state made out of a single massive particle at rest, we have demonstrated how  single-particle Goldstone boson states are instead replaced by particle-hole states. As the main result of this work, we have then shown that the boost Goldstone theorem is satisfied for Fermi liquids by the particle-hole continuum. Remarkably, this works out in the absence of a position-space Lagrangian describing the Fermi liquid's excitations. Instead, it relies entirely on Landau's original hypothesis that the Fermi liquid admits a quasiparticle description. It is worth noting that we have checked the theorem in the non-relativistic limit, where we have made use of the fact that the single-particle expectation values of the stress-energy tensor are given in terms of the mass and momentum of the quasiparticles.  We have also seen that away from the non-relativistic limit the relevant matrix elements receive higher order corrections due to forward scattering. 

Finally, let us emphasize that, to the best of our knowledge, this is the first example where a Goldstone theorem is obeyed not by single-particle states, but by a particle-hole continuum. Thanks to this peculiarity, Fermi liquids can evade the paradox of Sect.~\ref{heuristic}: there are gapless Goldstone {\em states}, but since these are not single-particle states, they are not interpolated by a local Goldstone {\em field}. The usual ideas of non-linearly realizing the broken symmetries through local Goldstone fields simply do not apply. It is not obvious what replaces them though: Rothstein and Shrivastava have studied what it means for the momentum-space Fermi liquid effective action to be consistent with spontaneously broken boosts \cite{Rothstein:2017twg,Rothstein:2017niq}, but it is not clear to us what the relationship is between their results and ours. For instance, they find that spontaneously broken boost invariance requires interactions in the IR to take the standard Landau form (back-to-back and forward scattering.) In comparison, we find that in the non-relativistic limit our Goldstone theorem is obeyed by the particle-hole continuum regardless of the structure of interactions.

We envision possible extensions of our results. For Fermi liquids, it would be interesting to go beyond the non-relativistic limit and see how the boost Goldstone theorem is obeyed in the relativistic case. In fact, by {\em requiring} that it be obeyed, we might learn some universal identities for relativistic Fermi liquid theory, which is certainly not as developed as the non-relativistic one.
For fermions at unitarity (see \emph{e.g.}~\cite{Nishida:2010tm} for a review), there are indications that the low-energy spectrum cannot be that of a Fermi liquid \cite{Rothstein:2017niq}. It would then be interesting to see how our boost Goldstone theorem is obeyed in that case. Perhaps related to this, the only technical difference between our Goldstone theorem compared to more standard ones is a derivative with respect to momentum, which stems directly from the broken generators' having an explicit dependence on the spacetime coordinates. Since the currents for conformal transformations also depend explicitly on coordinates, we wonder whether there could be systems in which conformal invariance is spontaneously broken yet there is no dilaton, with the associated Goldstone theorem being obeyed by a less conventional spectrum, akin to the particle-hole continuum of our Fermi liquid case.

\section*{Acknowledgements}
We would like to thank R.~Penco, F.~Piazza, R.~Rattazzi, I.~Rothstein, D.T.~Son,  F.~Strocchi, and G.~Villadoro for useful discussions and comments.
LA is supported by the European Research Council under the European Union's Seventh Framework Programme (FP7/2007-2013), ERC Grant agreement ADG 339140. AN is partially supported by the US DOE under grant number 
DE-SC011941 and by the Simons Foundation under award number 658906.

\appendix
\section*{Appendix}
\section{Spectral densities without Lorentz invariance}\label{spectral}

For a Poincar\'e invariant theory in a Poincar\'e invariant vacuum state $| \Omega \rangle$, the K\"allen-Lehmann representation conveniently expresses two-point functions of Lorentz covariant local operators in terms of free-field two-point functions and of spectral densities that are just functions of the intermediate states' invariant mass. For instance, the Feynman two-point function of a Lorentz scalar operator $O(x)$ is simply
\be
\tilde G_{OO}^F(p) = \int_0^\infty \frac{d\mu^2}{(2\pi)} \frac{i}{p^2 - \mu^2 + i \epsilon} \rho_{OO}(\mu^2) \; ,
\ee
where $\rho_{OO}(\mu^2)$ is a non-negative spectral density, normalized in such a way that for a canonically normalized, free scalar field
$\phi$ of mass $m$, the spectral density is $\rho_{\phi\phi}(\mu^2) = (2\pi) \delta(\mu^2 - m^2)$.

However, if the theory is not Lorentz invariant, or if the state on which we are computing correlation functions is not Lorentz invariant, the spectral representation of a generic correlation function will involve a spectral density that is a function separately of energy and momentum---the invariant mass combination plays no special role without Lorentz invariance. 

Without loss of generality, we can consider Wightman two-point functions, since all the other relevant ones (symmetric, anti-symmetric, retarded, advanced, $T$-ordered, anti-$T$-ordered, etc.) are suitable linear combinations of the Wigthman ones, possibly with $\theta$-function coefficients. 
For notational simplicity, let us restrict to hermitian operators. 
For two generic local operators $O_a$, $O_b$, the Wightman two-point function on a translationally invariant state $| \Omega \rangle$ is
\be
G_{ab}(x-y) \equiv \langle \Omega | O_a(x) O_b(y)| \Omega \rangle \; .
\ee
Inserting an orthogonal complete set of states as in \eqref{complete}, we have
\be
G_{ab}(x-y) = \int \frac{d^3 p}{(2\pi)^3} e^{i \vec p \cdot (\vec x- \vec y)} \sum_n e^{-i E_n(\vec p \, ) (t_x-t_y)}  {\cal O}_{a,n} (\vec p \, )   {\cal O}_{b,n} ^* (\vec p \, )\; ,
\ee
where the ${\cal O}$'s are the matrix elements
\be
{\cal O}_{a,n}(\vec p \, ) \equiv \langle \Omega | \, O_a(0) | n, \vec p \, \rangle \; .
\ee
One can then define the $O_a$-$O_b$ spectral density in a way analogous to the relativistic case, but without using Lorentz invariance:
\be \label{def rho}
\rho_{ab} (E, \vec p \, ) \equiv 2E \sum_n  {\cal O}_{a,n} (\vec p \, )   {\cal O}_{b,n} ^* (\vec p \, ) \times (2\pi) \delta(E - E_n(\vec p \, )) \; .
\ee
To guide the intuition, notice that if $|\Omega \rangle$ is the ground state of the system, and $O_a = O_b = \phi$, with $\phi$ being a canonically 
normalized, free, real scalar field that creates and annihilates single-particle states with generic dispersion relation $E(\vec p \,)$,
\be \label{phi}
\phi(x) = \int \frac{d^3 p}{(2 \pi)^3 } \frac{1}{\sqrt{2 E(\vec p \, )}} \big[ a_{\vec p }  \, e^{-i E(\vec p \,) t} e^{i \vec p \cdot \vec x}+ a^\dagger_{\vec p}  \,  e^{+i E(\vec p \, ) t} e^{-i \vec p \cdot \vec x}\big] \; ,
\ee
the spectral density reduces simply to $\rho_{\phi\phi} (E, \vec p \, ) =   (2\pi) \delta(E - E(\vec p \, ))$, which is what we want the spectral density to be in this case.

Going back to the general case, using  \eqref{def rho}, our two-point function becomes
\be
G_{ab}(x-y) = \int \frac{d^3 p}{(2\pi)^3} \frac{dE}{(2\pi)} \frac{1}{2E} e^{-i E (t_x-t_y)}  e^{i \vec p \cdot (\vec x- \vec y)} \rho_{ab} (E, \vec p \, ) \; ,
\ee
or, in Fourier transform,
\begin{align}\label{spectral1}
\tilde G_{ab} (\vec p \,, \omega) & = \int \frac{dE}{(2\pi)} \,  \frac{1}{2E} (2 \pi) \delta(\omega - E) \, \rho_{ab}(E, \vec p)  \\
& = \frac{1}{2 \omega} \rho_{ab}(\omega, \vec p \, ) 
\; .
\end{align}

The final result is of course very simple and convenient, but it only holds for Wightman two-point functions. The intermediate integral expression \eqref{spectral1} is conceptually more useful, in that it highlights that our two-point function is, like in the relativistic case, a convolution of a free-field two-point function with a spectral density. Indeed, for the free scalar of eq.~\eqref{phi}, the Wightman two-point function is\footnote{As a check, notice that for a relativistic dispersion relation, $E(\vec p \,) = \sqrt{\vec p \,^2 + m^2}$, eq.~\eqref{Wightman free} can be written in a manifestly covariant form as
\be
\tilde G_{\phi\phi} (\vec p \,, \omega) =  \theta(\omega) \, (2 \pi) \delta(p^2-m^2) \; ,
\ee
which is the correct Wightman two-point function for a free relativistic scalar field.}
\be \label{Wightman free}
\tilde G_{\phi\phi} (\vec p \,, \omega) =  \frac{1}{2 E(\vec p \,)} (2 \pi) \delta(\omega - E( \vec p \, ))
\ee
So, a general two-point function, with general ordering of the operators, will be
\be
\tilde G^{\rm gen}_{ab} (\vec p \,, \omega)  = \int \frac{d E}{(2\pi)} \,  \tilde G^{\rm gen}_{\rm free} (\vec p \, , \omega; E) \, \rho_{ab}(E, \vec p)  \; ,
\ee
where $\rho_{ab}(E, \vec p \,)$ is the same spectral density defined above, and $G^{\rm gen}_{\rm free} (\vec p \, , \omega; E) $ is the corresponding two-point function for a canonically normalized free scalar field that, at momentum $\vec p$, creates and annihilates excitations of energy $E$.

The properties discussed above are all we need for our purposes, but there is of course a vast literature on the subject of spectral functions for non-relativistic situations, including for systems at finite temperature. See \emph{e.g.}~\cite{Hartnoll:2009sz} and references therein.

\section{A quick analysis of the Fermi liquid case}\label{quick}
To check if our Goldstone theorem is obeyed by particle-hole states in a Fermi liquid, we need to compute the matrix elements
\be
\langle {\rm FL} | T^{\mu\nu} (0) \, c^\dagger_1 c_2 | {\rm FL} \rangle \; , \label{ph}
\ee
where we are using the shorthand notation $c_a \equiv c_{\vec p_a}^{s_a}$. In particular, we need these matrix elements in the limit $\vec p \equiv \vec p_1 - \vec p_2 \to 0$, but with  $p_1 > p_F$ (particle) and  $p_2<p_F$ (hole). 

We want to relate these to the particle-particle matrix elements
\be
\langle {\rm FL} | c_2 \, T^{\mu\nu} (0) \, c^\dagger_1 | {\rm FL} \rangle \; , \label{pp}
\ee
still in the limit $\vec p \equiv \vec p_1 - \vec p_2 \to 0$, but now with  $p_1, p_2 > p_F$ (particle, particle). The reason we want to do so is that these are easy to compute. In particular, for the entries relevant for our theorem in the non-relativistic limit we have, {\em by definition},
\be \label{relevant for theorem}
\langle {\rm FL} | c_2 \, T^{00} (0) \, c^\dagger_1 | {\rm FL} \rangle \to m  \; ,\qquad \langle {\rm FL} | c_2 \, T^{0i} (0) \, c^\dagger_1 | {\rm FL} \rangle \to p_2^i 
\ee
(Cf.~Section {\ref{sec:massive_part}).

In order to relate \eqref{ph} to \eqref{pp}, we need to move $c_2$ all the way to left. Clearly, we want to use the fact that $c_2$ has simple anticommutation relations with the other  $c$'s and $c^\dagger$'s. This means that whenever we need to move $c_2$ through an operator with an odd number of $c$'s and $c^\dagger$'s, we want to use an anticommutator, whereas whenever we need to move $c_2$ through an operator with an even number of $c$'s and $c^\dagger$'s, we want to use a commutator.\footnote{These simple rules of thumb follow from using recursively $[AB, C] = A\{B,C\}-\{A,C\}B$ and $\{AB,C\}=A\{B,C\} - [A,C]B$ (or variations thereof), which are valid for any three operators $A, B, C$.} 
Notice that  in both \eqref{ph} and \eqref{pp} we are keeping $\vec p_1 \neq \vec p_2$, and so $c_1^\dagger$ and $c_2$ anticommute. So, we
write
\be\label{D4}
T^{\mu\nu} (0) \, c^\dagger_1 c_2 = - c_2 \, T^{\mu\nu} (0) \, c^\dagger_1 - \big[ T^{\mu\nu} (0), c_2 \big] c_1^\dagger\,, \qquad\qquad \big( \vec p_1 \neq \vec p_2 \big) \; ,
\ee
where we used the fact that
the stress-energy tensor operator, being a bosonic operator, must be a sum of terms with an even number of $c$'s and $c^\dagger$'s.  

It is more useful to write the above identity as
\be
T^{\mu\nu} (0) \, c^\dagger_1 c_2 + c_2 \, T^{\mu\nu} (0) \, c^\dagger_1 = - \big[ T^{\mu\nu} (0), c_2 \big] c_1^\dagger\,, \qquad\qquad \big( \vec p_1 \neq \vec p_2 \big) \; . \label{identity}
\ee
Now imagine evaluating both sides of this equation on $|{\rm FL}\rangle$, keeping $\vec p_1$ always above the Fermi surface, but moving $\vec p_2$ from just below  to just above the Fermi surface: on the l.h.s., the first term goes from being nonzero ($p_2<p_F$) to being zero ($p_2>p_F$), while the second term goes from being zero ($p_2<p_F$) to being non-zero ($p_2>p_F$). On the other hand, the r.h.s.~is continuous when $\vec p_2$ crosses the Fermi surface, because $\big[ T^{\mu\nu} (0), c_2 \big]$ involves anticommutators of $c_2$ with other $c$'s and $c^\dagger$'s, which remove $c_2$---the only source of discontinuity. Note that $c_2$ does not drop out if instead we use the anticommutator $\{T^{\mu\nu}(0),c_2\}$ (see, for example, \eqref{commutator_p2}) --- the reason why expressing \eqref{D4} in terms of a commutator is more convenient in this case. 

More concretely, parametrizing $T^{\mu\nu} (0)$ as a sum of terms schematically of the form (as done in Section~\ref{sec:interactions})
\be
\int_{k_1, \dots k_n, q_1, \dots q_n} F^{\mu\nu}\big(\vec k_1, \dots, \vec k_n;\vec q_1,\dots \vec q_n \big) c^\dagger_{k_1} \dots c^\dagger_{k_n} c_{q_1} \dots c_{q_n}\,,
\ee
(we use the shorthand notation  $\int_{\vec p} \equiv \int \frac{d^3 p}{(2\pi)^3}$ and a sum over spin labels is understood), when we take the commutator with $c_2$ we are left with terms of the form 
\be
\int_{k_2 \dots k_n, q_1, \dots q_n} F^{\mu\nu}\big(\vec p_2, \vec k_2, \dots, \vec k_n;\vec q_1,\dots \vec q_n \big) c^\dagger_{k_2} \dots c^\dagger_{k_n} c_{q_1} \dots c_{q_n} \; .
\ee
The only dependence on $\vec p_2$ is now inside $F^{\mu\nu}$, but this function is continuous when its momentum arguments cross the Fermi surface,  because  moving a quasiparticle from slightly below  to slightly above the Fermi surface does not change its physical properties nor its interactions with the rest of the Fermi liquid in a discontinuous manner.

In conclusion, since the r.h.s.~of \eqref{identity}  evaluated on $|{\rm FL} \rangle$ is continuous for $\vec p_2$ crossing the Fermi surface, we have
\be \label{two limits}
\lim_{p_2 \to p_{F-}}\langle {\rm FL} | T^{\mu\nu} (0) \, c^\dagger_1 c_2 | {\rm FL} \rangle = \lim_{p_2 \to p_{F+}} 
\langle {\rm FL} | c_2 \, T^{\mu\nu} (0) \, c^\dagger_1 | {\rm FL} \rangle \; .
\ee
And, so, given \eqref{relevant for theorem} above, for our theorem we have
\be
\lim_{\vec p \to 0} \Big( {\cal T}^{00} {\cal T}^{0j} {}^* + {\rm c.c.} \big) = 2 m \, p_2^j \; .
\ee
As shown in Appendix \ref{app:integral}, integrating this over the phase space of particle-hole states with total momentum $\vec p$, yields, to first order in $\vec p$,
\be
\frac{1}{6\pi^2} m p_{\rm F}^3 \, \vec p \; ,
\ee
which obeys our theorem, once we take into account an extra factor of 2 coming from the sum over spin states, and the fact that in (non-relativistic) Fermi liquid theory the mass density is  $\frac{1}{3 \pi^2} m p_{\rm F}^3$ (again including a factor of 2 coming from the spin states) \cite{Lifshitz}.

Notice that a minor variation of \eqref{identity} applies to the bosonic case discussed in Sects.~\ref{free bosons}, \ref{symmetry free bosons}, \ref{symmetry interacting bosons}:
\be
T^{\mu\nu} (0) \, a^\dagger_1 a_2 - a_2 \, T^{\mu\nu} (0) \, a^\dagger_1 = \big[ T^{\mu\nu} (0), a_2 \big] a_1^\dagger, \qquad\qquad \big( \vec p_1 \neq \vec p_2 \big) \; . \label{bosons} 
\ee
However, the crucial difference with the fermionic case is that when we evaluate both sides on a Fermi liquid-like state, the second term on the l.h.s.~now is nonzero even for $\vec p_2$ {\em below} the Fermi surface: there is no Pauli exclusion principle. As a result, in general there is no relationship similar to \eqref{two limits} for bosonic Fermi liquid-like states.  The exception is the free boson case discussed in sects.~\ref{free bosons}, \ref{symmetry free bosons}, where the expectation values on $|{\rm FL} \rangle$ of the r.h.s.~of \eqref{bosons} and of the second term on the l.h.s.~happen to be proportional to each other.

\section{The integral}\label{app:integral}
In this Section we perform the integral given in Eq. \eqref{integral} for our particle--hole excited state around the Fermi liquid ground state. To do so we first go to a spherical coordinate system with $z$-axis aligned along $\vec p$. The integration variables are then the absolute value $p_2\equiv |\vec p_2|$ and the two angles $\varphi$, $\theta'$. Note that $\theta'=\pi-\theta$, where $\theta$ is the angle between $\vec p$ and $\vec p_2$. The corresponding setup is shown in Fig.~\ref{fig:coordinates}. We thus have:
\be
\int\frac{d^3p_2}{(2\pi)^3}\left(2p_2^i+p^i\right)=\frac{1}{(2\pi)^3}\int d\varphi \,d\theta' dp_2\,p_2^2\sin\theta'(2p_2^i+p^i)\,.
\ee

\begin{figure}[h]
    \centering
\includegraphics[width=6.5cm]{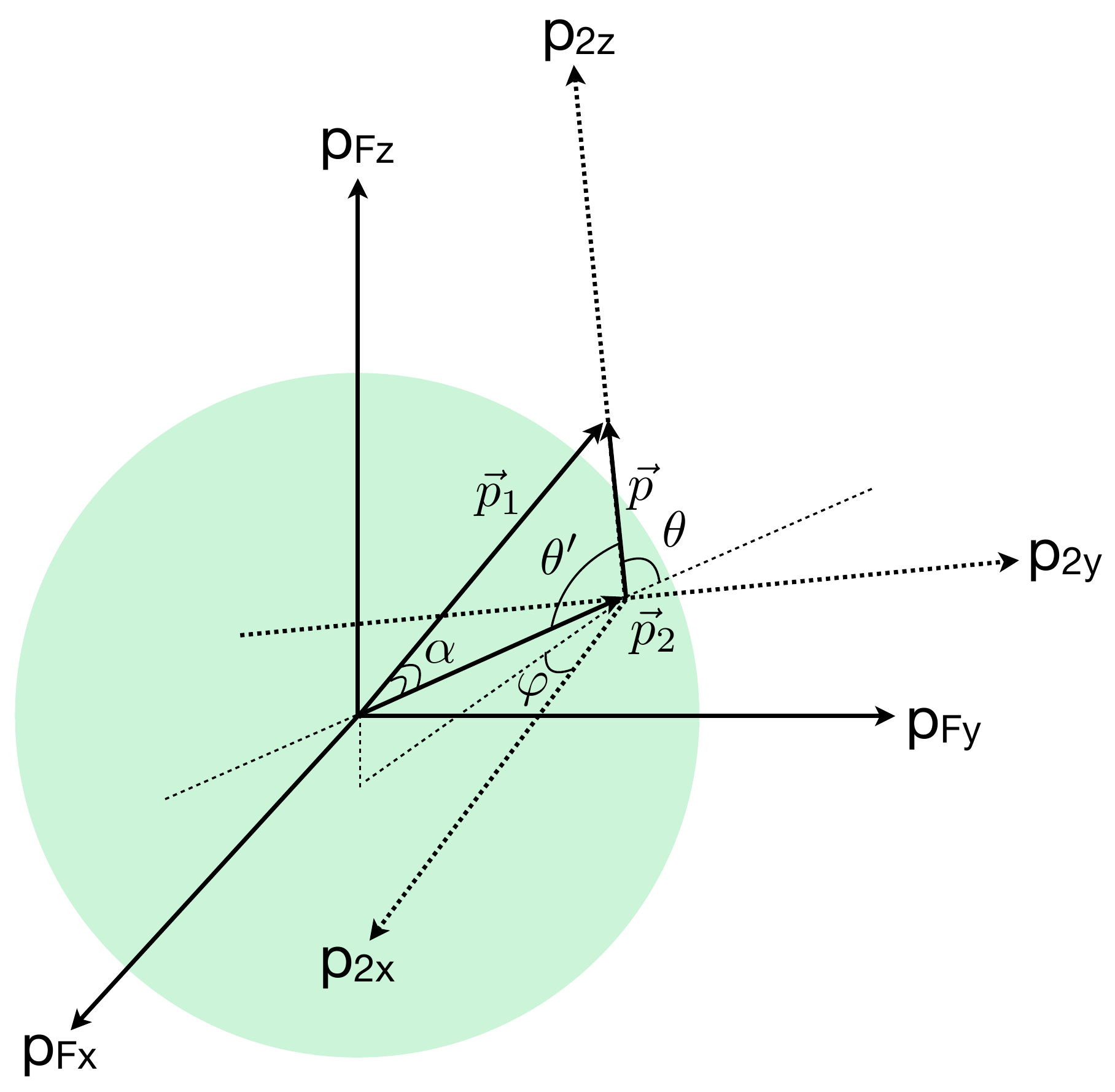}
\caption{The coordinate system used in performing the integral \eqref{integral}.}
\label{fig:coordinates}
\end{figure}

Let us establish the integration limits for a given vector $\vec p$, independent from $\vec p_2$. We shall make use of the fact that $\vec p_1=\vec p+\vec p_2$ and paramterize the absolute values as
\be
p_1\equiv |\vec p_1|=p_{\rm F}+\delta p_1\,,\quad p_2\equiv |\vec p_2|=p_{\rm F}-\delta p_2\,,\quad 0\leq\frac{\delta p_1}{p_{\rm F}}\,,\frac{\delta p_2}{p_{\rm F}}\ll 1\,,
\ee
where the last condition arises due to the fact that it is only in the region close to the Fermi surface that the quasiparticles can be treated as weakly interacting. 
The range of integration of $\varphi$ is not affected by these constraints and remains $\varphi\in[0,2\pi]$. However, from the relationship $|\vec p_1|^2=(\vec p+\vec p_2)^2$ and demanding that $\delta p_1\geq 0$ there is an upper bound on the absolute value of $\delta p_2$ leading to the range of integration:
\be\label{dp2max}
0\leq \delta p_2\leq\delta p_{2,\text{max}}\equiv \frac{-p\cos\theta'+\frac{p^2}{2p_{\rm F}}}{1-\frac{p}{p_{\rm F}}\cos\theta'}+\mathcal O(\delta p_i^2) \,.
\ee
Note that we have not constrained the absolute value of $\vec p/p_{\rm F}$. Although we shall take the limit $\vec p\to 0$ in order to prove the Goldstone theorem, for the sake of completeness we choose not to impose this constraint here. The reason is that although Landau's theory does say that the system of particles and holes is weakly interacting only in the vicinity of the Fermi surface, it imposes no condition on the total momentum of the system $\vec p$. In other words, while we do demand that $\delta p_i/p_{\rm F}\ll 1$ the total momentum can take the values in the range $p\in [0,2p_{\rm F}]$. 

It remains to find the range of integration for $\theta'$ for a given $\vec p$. Obviously, the minimal value of $\theta'$ occurs when the particle and hole are almost antipodal and $p\approx 2p_{\rm F}$ in which case $\theta' \to 0$. To find its maximal allowed value it turns out to be useful to express $\cos \alpha$ (the angle $\alpha$ is defined in Fig.~\ref{fig:coordinates}) by making use of the cosine theorem for $\alpha$. This leads to
\be
\cos \alpha=1-\frac{p^2}{2p_{\rm F}^2}+\mathcal O(\delta p_i)\,.
\ee
The angles $\theta'$ and $\alpha$ can in turn be related by 
\be
\cos \theta'=\frac{1}{p}(p_2-p_1\cos\alpha)\,\qquad\Rightarrow\qquad \cos\theta'_{\text{max}}=\frac{p}{2p_{\rm F}}+\mathcal O(\delta p_i)\,.
\ee
The integral above then becomes
\be\label{C6}
\int\frac{d^3p_2}{(2\pi)^3}\left(2p_2^i+p^i\right)=\frac{1}{(2\pi)^3}\int_0^{2\pi} d\varphi \,\int_{\frac{p}{2p_{\rm F}}}^1d\cos \theta' \int _{0}^{\delta p_{2,\text{max}}}d\delta p_2\,(p_{\rm F}-\delta p_2)^2(2p_2^i+p^i)
\ee
where we have changed the integration variable from $p_2$ to $\delta p_2$ and  $\delta p_{2,\text{max}}$ is defined in \eqref{dp2max}. We note that in the small $\vec p$ limit the integration \emph{measure} becomes
\be\label{integral_measure}
\lim_{\vec p\to 0} \int\frac{d^3p_2}{(2\pi)^3}=\frac{p_{\rm F}^2}{(2\pi)^3}\int_0^{2\pi} d\varphi \,\int_{0}^1d\cos \theta' \int _{0}^{-p\cos\theta'}d\delta p_2+\mathcal O(p^2)\,.
\ee
Hence, due to the upper integration limit for $\delta p_2$, this means that the integration measure itself is already linear in $p$ and only the last term in \eqref{C6} needs to be considered. As the last step we note that only the $z$ component  of $p_2^i$ survives after the integration over $\varphi$ is performed. In small $p$ limit we thus have
\be\label{integral_result}
\begin{split}
\lim_{\vec p\to 0}\int\frac{d^3p_2}{(2\pi)^3}\left(2\vec p_2+\vec p\right)&=-\frac{1}{4\pi^2}p_{\rm F}^2p\int_{\frac{p}{2p_{\rm F}}}^1d\cos \theta' \cos\theta'(2p_2\hat p+\mathcal O(p))\\
&=\frac{1}{6\pi^2}p_{\rm F}^3\vec p+\mathcal O(p^2)\,,
\end{split}
\ee
where $\hat p$ denotes the unit vector in the direction of $\vec p$ (which coincides with the $z$ axis).

\section{Stress--energy tensor for free fermions}\label{app:free_fermions}
As we are interested in the non-relativistic limit, we consider the solutions to the Dirac equations in Dirac representation where the gamma matrices are defined as
\be
\gamma^i=\begin{pmatrix} 0&\sigma^i\\-\sigma^i&0\end{pmatrix}\,,\qquad
\gamma^0=\begin{pmatrix}\mathbb 1&0\\0&-\mathbb 1\end{pmatrix}\,,\qquad
\gamma^5=\begin{pmatrix}0&\mathbb 1\\\mathbb 1&0\end{pmatrix}\,.
\ee 
The solutions for the Dirac spinors in \eqref{modes_ferm} are then up to first order in the non-relativistic limit:
\be
u^s_{\vec p}=\sqrt{2m}\begin{pmatrix}\xi^s_p\\[0.5em]\frac{\vec\sigma\cdot\vec p}{2m}\xi^s_p\end{pmatrix}\,,\qquad 
v^s_{\vec p}=\sqrt{2m}\begin{pmatrix}\frac{\vec\sigma\cdot\vec p}{2m}\chi^s_p\\[0.5em]\chi^s_p\end{pmatrix}\,,
\ee
where 
\be
\xi^1_p = \begin{pmatrix}1\\0\end{pmatrix}\,,\qquad 
\xi^2_p=\begin{pmatrix}0\\1\end{pmatrix}\,,\qquad
\chi^s_p=\frac{\vec\sigma\cdot\vec p}{2m}\xi^s_p\,.
\ee
We then use the non-relativistic limit relations
\be
u^{s\dagger}_{\vec p}u^{s'}_{\vec k}=v^{s\dagger}_{\vec p}v^{s'}_{\vec k}=2m\delta^{ss'}\,,\qquad u^{s\dagger}_{\vec p}v^{s'}_{\vec k}=v^{s\dagger}_{\vec p}u^{s'}_{\vec k}=0
\ee
in order to find the $T^{00}(0)$ components. Similarly, we derive the following relationships
\be
\begin{split}
&u^{s\dagger}_{\vec p}\gamma^0\gamma^iu^{s'}_{\vec k}=\xi^{s\dagger}\left(\sigma^i\sigma^j k^j+p^j\sigma^j\sigma^i\right)\xi^{s'}\,,\\
&v^{s\dagger}_{\vec p}\gamma^0\gamma^iv^{s'}_{\vec k}=\xi^{s\dagger}\left(\sigma^i\sigma^j(k^j-p^j)+2\delta^{ij}p^j\right)\xi^{s'}\,,\\
&v^{s\dagger}_{\vec p}\gamma^0\gamma^iu^{s'}_{\vec k}=u^{s\dagger}_{\vec p}\gamma^0\gamma^iv^{s'}_{\vec k}=2m\xi^{s\dagger}\sigma^i\xi^{s'}\,,
\end{split}
\ee
where we have used the fact that in the non-relativistic limit $\xi^s_p$ does not depend on $p$ and have dropped the subscript. These allow one to find the $T^{0i}(0)$ components. The final expressions are given in Eqns.~\eqref{T00_ferm}, \eqref{T0i_ferm} in the main text.

\section{Matrix elements}\label{sec:elements}
Here we evaluate the matrix elements $\langle \mathbb 1_{\vec p_2}^{s_2}|T^{\mu\nu}_{\rm free}(0)|\mathbb 1_{\vec p_1}^{s_1}\rangle$ and $\left\langle \bar{\mathbb 1}_{\vec p_2}^{s_2}\right|T^{\mu\nu}_{\rm free}(0)\left|\bar{\mathbb 1}_{\vec p_1}^{s_1}\right\rangle$ for  various relevant configurations of momenta $\vec p_1$ and $\vec p_2$: 

\begin{itemize}
\item{$|\vec p_1|>p_F$ and $|\vec p_2|>p_F$} 

When both momenta are above the Fermi surface, the states $\langle\mathbb 1^{s_2}_{\vec p_2}|\equiv  \langle\text{FL}|b^{s_2}_{\vec p_1}$ and $|\mathbb 1^{s_1}_{\vec p_1}\rangle\equiv b^{\dagger s_{1}}_{\vec p_1}|\text{FL}\rangle$ both correspond to single \emph{particle} states. In this case, we find
\be
\begin{split}
&\left\langle \mathbb 1^{s_2}_{\vec p_2}\right|b^{s\,\dagger}_{\vec k}b^{s'}_{\vec q}\left|\mathbb 1^{s_1}_{\vec p_1}\right\rangle\equiv \left\langle \text{FL}\right|b^{s_2}_{\vec p_2}b^{s\,\dagger}_{\vec k}b^{s'}_{\vec q}b^{s_1\dagger}_{\vec p_1}\left|\text{FL}\right\rangle 
 = \delta_{p_1q}\delta_{p_2k}+\left.\delta_{p_1p_2}\delta_{qk}\right|_{|\vec k|,|\vec q|\leq p_F}
\end{split}
\ee
for both bosonic and fermionic creation and annihilation operators. Here and henceforth, we use the shorthand notations $\delta_{p_1p_2}\equiv(2\pi)^3\delta^{s_1s_2}\delta^{(3)}(\vec p_1-\vec p_2)$ etc. 
 This allows one to evaluate 
\be\label{Tmn_single_particle}
\left\langle \mathbb 1^{s_2}_{\vec p_2}\right|T^{\mu\nu}_{\rm free}(0)\left|\mathbb 1^{s_1}_{\vec p_1}\right\rangle=F^{\mu\nu}_{s_2s_1}(p_2,p_1)+\delta_{p_1p_2}\langle \text{FL}|T^{\mu\nu}_{\rm free}(0) |\text{FL}\rangle\,.
\ee
Comparing this with the non-relativistic single-particle relations \eqref{match_00} and \eqref{match_0i} we obtain:\footnote{One could in fact be more precise when defining the function $F^{\mu\nu}_{ss'}(k,p)$. In particular, we can split
\be
\begin{split}\label{ansatz2}
F^{00}_{ss'}(k,q)&=f^0(k,q)\delta^{ss'}+ g^i(k,q) \sigma_i^{ss'}\,,\\
F^{0i}_{ss'}(k,q)&=f^i(k,q)\delta ^{ss'}+ g(k,q) \sigma_i^{ss'}+\varepsilon_{ijk}h^j(k,q)\sigma_k^{ss'}\,.
\end{split}
\ee
In the above expressions, $\sigma_i$ are the Pauli matrices and we have accounted for the different spins by expanding in a basis of Hermitian $2\times 2$ matrices. From \eqref{Fs} we can then read off: $f^0(p,p)=m$, $g^i(p,p)=0$ and $f^i(p,p)=p^i$, $g(p,p)=0$, $ h^i(p,p)=0$.}
\be\label{Fs}
F^{00}_{ss}(p,p)=m\,,\qquad F^{0i}_{ss}(p,p)=p^i\,,\qquad\text{for }|\vec p \, |> p_F\,.
\ee

\item{$|\vec p_1|\leq p_F$ and $|\vec p_2| \leq p_F$}

The situation is slightly more complex when both momenta are below the Fermi surface, since the bosonic and fermionic cases require a separate treatment. In this case the states $\langle\bar{\mathbb 1}^{s_2}_{\vec p_2}|\equiv \langle\text{FL}|b^{s_2\dagger}_{\vec p_2}$ and $|\bar{\mathbb 1}^{s_1}_{\vec p_1}\rangle\equiv b^{s_{1}}_{\vec p_1}|\text{FL}\rangle$ correspond to single \emph{hole} states. To evaluate the matrix elements for the single hole states we first find for fermions:
\be
\begin{split}
&\left\langle \bar{\mathbb 1}_{\vec p_2}^{s_2}\right|c^{s \dagger}_{\vec k}c^{s'}_{\vec q}\left|\bar{\mathbb 1}_{\vec p_1}^{s_1}\right\rangle\equiv\left\langle \text{FL}\right|c^{s_2 \dagger }_{\vec p_2}c^{s\,\dagger}_{\vec k}c^{s'}_{\vec q}c^{s_1}_{\vec p_1}\left|\text{FL}\right\rangle=-\delta_{p_1k}\delta_{p_2q}+\left.\delta_{p_1p_2}\delta_{qk}\right.\,,
\end{split}
\ee
where we have made use of the fact that the above matrix elements vanish unless also $|\vec q|, |\vec k|\leq p_{\rm F}$ and that for a fermionic Fermi liquid $c^{s_1 \dagger }_{\vec p_1}|\text{FL}\rangle=0$ for any momentum $|\vec p_1|\leq p_{\rm F}$. This leads to
\be\label{single_hole}
\left\langle \bar{\mathbb 1}_{\vec p_2}^{s_2}\right|T^{\mu\nu}_{\rm free}(0)\left|\bar{\mathbb 1}_{\vec p_1}^{s_1}\right\rangle=-F^{\mu\nu}_{s_1s_2}(p_1,p_2)+\delta_{p_1p_2}\langle \text{FL}|T^{\mu\nu}_{\rm free}(0) |\text{FL}\rangle\,.
\ee
Note that the order in the arguments of the first term is different from the result for single \emph{particle} states in \eqref{Tmn_single_particle}. Using as before that $\rho_{\rm FL}=\langle \text{FL}|T^{\mu\nu}_{\rm free}(0) |\text{FL}\rangle$ and matching with the requirements \eqref{match00h} and \eqref{match0ih} for the non-relativistic case we obtain that for fermions:
\be\label{Fs2}
F^{00}_{ss}(p,p)=m\,,\qquad F^{0i}_{ss}(p,p)=p^i\,,\qquad |\vec p|\leq p_{\rm F}\,.
\ee

For bosons $a^{\dagger}_{\vec p_1}|\text{FL}\rangle\neq0$ (while $a_{\vec p_1}|\text{FL}\rangle=0$ still holds) and thus the matrix elements $\left\langle \text{FL}\right|a^{\dagger}_{\vec p_2}a^{\dagger}_{\vec k}\,a_{\vec q}\,a_{\vec p_1}\left|\text{FL}\right\rangle$ have to be evaluated explicitly. It is instructive to first evaluate it between two-particle states like $|a,b\,\rangle \equiv a^{\dagger}_{\vec q_a}a^{\dagger}_{\vec q_b}|0\rangle$. This gives $\langle c,d\,|a^{\dagger}_{\vec p_2}a^{\dagger}_{\vec k}\,a_{\vec q}\,a_{\vec p_1}|a,b\,\rangle=(\delta_{q_ck}\delta_{q_dp_2}+\delta_{q_cp_2}\delta_{q_dk})(\delta_{q_ap_1}\delta_{qq_b}+\delta_{qq_a}\delta_{q_bp_1})$. We can then recall that the Fermi liquid state $|\rm FL\rangle$ was defined in \eqref{FL_boson} as a tensor product of all single particle states with momenta below the Fermi surface. Importantly, it is a product of single particle states of \emph{distinct} momenta. In the above example this would mean that only one of the $\delta$'s in each bracket can be satisfied simultaneously, \emph{i.e.} if $ \vec q_a=\vec q$ this means that $\vec q_a\neq \vec p_1$ etc. For a Fermi liquid ground state and for $|\vec p_1|,|\vec p_2|,|\vec q|,|\vec k|\leq p_{\rm F}$ this means
\be\label{boson_normalization}
\begin{split}
\left\langle \text{FL}\right|a^{\dagger }_{\vec p_2}a^{\dagger}_{\vec k}\,a_{\vec q}\,a_{\vec p_1}\left|\text{FL}\right\rangle=\delta_{p_1k}\delta_{p_2q}+\left.\delta_{p_1p_2}\delta_{qk}\right.-\frac{2}{V}\delta_{p_2k}\delta_{kq}\delta_{qp_1}\,,
\end{split}
\ee
where we have used $\langle\text{FL}|\text{FL}\rangle=1$. The first two terms above give the correct result when the momenta in the creation and annihilation `pairs' are different, \emph{i.e.} $ \vec p_2\neq \vec k$, $\vec p_1\neq \vec q$, while the last term accounts for the special case when all the momenta are equal. Its normalization is determined by demanding that $\langle\text{FL}|a^\dagger_{\vec p}\, a^\dagger_{\vec p}\, a_{\vec p}\,a_{\vec p}|\text{FL}\rangle =0$ due to the property $a_{\vec p}\,a_{\vec p}|\text{FL}\rangle = 0$ for $|\vec p|\leq p_{\rm F}$. The above result thus gives
\be\label{hole_hole}
\left\langle \bar{\mathbb 1}_{\vec p_2}\right|T^{\mu\nu}_{\rm free}(0)\left|\bar{\mathbb 1}_{\vec p_1}\right\rangle=F^{\mu\nu}(p_1,p_2)-2F^{\mu\nu}(p_1,p_2)\frac{\delta^{(3)}(\vec p_1-\vec p_2)}{\delta^{(3)}(0)}+\delta_{p_1p_2}\langle \text{FL}|T^{\mu\nu}_{\rm free}(0) |\text{FL}\rangle\,.
\ee
Comparing with \eqref{match_00} and \eqref{match_0i} in a continuous way gives:
\be
\begin{split}
F^{00}(p,p)=m\,,\qquad F^{0i}(p,p)=p^i\,,\qquad |\vec p|\leq p_{\rm F}\,,
\end{split}
\ee
as for the case of fermions in \eqref{Fs2}. Together with \eqref{Fs} this means that the functions $F^{\mu\nu}(p,p)$ are continuous at the Fermi surface.

\item{$|\vec p_1|> p_F$ and $|\vec p_2| \leq p_F$}

At last, let us consider the particle-hole momentum configuration appearing in \eqref{state_relation}.  We find that $\left\langle {\mathbb 1}_{\vec p_2}^{s_2}\right|c^{s\,\dagger}_{\vec k}c^{s'}_{\vec q}\left|{\mathbb 1}_{\vec p_1}^{s_1}\right\rangle\equiv\left\langle \text{FL}\right|c^{s_2}_{\vec p_2}c^{s\,\dagger}_{\vec k}c^{s'}_{\vec q}c^{s_1\dagger}_{\vec p_1}\left|\text{FL}\right\rangle=0$ for fermions, due to property \eqref{vanishFL}. Instead, for bosons we find
\be\label{bosonic_elem}
\begin{split}
&\left\langle {\mathbb 1}_{\vec p_2}^{s_2}\right|a^{s\,\dagger}_{\vec k}a^{s'}_{\vec q}\left|{\mathbb 1}_{\vec p_1}^{s_1}\right\rangle\equiv \left\langle \text{FL}\right|a^{s_2}_{\vec p_2}a^{s\,\dagger}_{\vec k}a^{s'}_{\vec q}a^{s_1\dagger}_{\vec p_1}\left|\text{FL}\right\rangle=2\delta_{p_1q}\delta_{p_2k}
+\delta_{p_1p_2}\delta_{qk}\,.
\end{split}
\ee
Note the factor of $2$ in the first term. This leads to
\be\label{part_part}
\left\langle \mathbb 1^{s_2}_{\vec p_2}\right|T^{\mu\nu}_{\rm free}(0)\left|\mathbb 1^{s_1}_{\vec p_1}\right\rangle=2F^{\mu\nu}_{s_2s_1}(p_2,p_1)+\delta_{p_1p_2}\langle \text{FL}|T^{\mu\nu}_{\rm free}(0) |\text{FL}\rangle\,.
\ee
\end{itemize}

\bibliographystyle{apsrev4-1}
\bibliography{references}

\end{document}